\documentclass[a4paper,manyauthors,nocleardouble,COMPASS]{cernphprep}
\usepackage[utf8]{inputenc}
\usepackage{amssymb}
\usepackage{amsmath}
\usepackage{lineno} 
\usepackage{xspace}
\usepackage{graphicx}
\usepackage{caption}
\usepackage{subcaption}
\usepackage{color}
\usepackage{float}
\usepackage{datetime}
\usepackage{cite}
\usepackage{hyperref}

\newcommand{\varQtwo}[1]{$Q^{2}$\xspace}

\begin{document}

\begin{titlepage}
\PHnumber{2017--xxx}
\PHdate{January 2017}

\everymath{\displaystyle}

\title{First measurement of the Sivers asymmetry for gluons from SIDIS data}

\begin{abstract}
The Sivers function describes the correlation between the transverse spin of a nucleon and the transverse motion of its partons. It was extracted from measurements of the azimuthal asymmetry of hadrons produced in semi-inclusive deep inelastic scattering of leptons off transversely polarised nucleon targets, and it turned out to be non-zero for quarks. In this letter the evaluation of the Sivers asymmetry for gluons in the same process is presented. The analysis method is based on a Monte Carlo simulation that includes three hard processes: photon-gluon fusion, QCD Compton scattering and leading-order virtual-photon absorption process. The Sivers asymmetries of the three processes are simultaneously extracted using the LEPTO event generator and a neural network approach. The method is applied to samples of events containing at least two hadrons with large transverse momentum  from the COMPASS data taken with a 160~GeV/$c$ muon beam scattered off transversely polarised deuterons and protons. With a significance of more than two standard deviations a negative value is obtained for the gluon Sivers asymmetry. The result of a similar analysis for a Collins-like asymmetry for gluons is consistent with zero. 
\end{abstract}

\vspace*{60pt}
Keywords: deep inelastic scattering, gluon, Sivers, TMD, PDF.

\vfill
\Submitted{(to be submitted to Phys.\ Lett.\ B)}

%
%
\section*{The COMPASS Collaboration}
\label{app:collab}
\renewcommand\labelenumi{\textsuperscript{\theenumi}~}
\renewcommand\theenumi{\arabic{enumi}}
\begin{flushleft}
C.~Adolph\Irefn{erlangen},
M.~Aghasyan\Irefn{triest_i},
R.~Akhunzyanov\Irefn{dubna}, 
M.G.~Alexeev\Irefn{turin_u},
G.D.~Alexeev\Irefn{dubna}, 
A.~Amoroso\Irefnn{turin_u}{turin_i},
V.~Andrieux\Irefnn{illinois}{saclay},
N.V.~Anfimov\Irefn{dubna}, 
V.~Anosov\Irefn{dubna}, 
A.~Antoshkin\Irefn{dubna}
% ...
\end{flushleft}
%
%
\begin{Authlist}
\item \Idef{turin_p}{University of Eastern Piedmont, 15100 Alessandria, Italy}
\item \Idef{aveiro}{University of Aveiro, Dept.\ of Physics, 3810-193 Aveiro, Portugal}
\item \Idef{bochum}{Universit\"at Bochum, Institut f\"ur Experimentalphysik, 44780 Bochum, Germany\Arefs{l}\Arefs{s}}
\item \Idef{bonniskp}{Universit\"at Bonn, Helmholtz-Institut f\"ur  Strahlen- und Kernphysik, 53115 Bonn, Germany\Arefs{l}}
\item \Idef{bonnpi}{Universit\"at Bonn, Physikalisches Institut, 53115 Bonn, Germany\Arefs{l}}
\item \Idef{brno}{Institute of Scientific Instruments, AS CR, 61264 Brno, Czech Republic\Arefs{m}}
\item \Idef{calcutta}{Matrivani Institute of Experimental Research \& Education, Calcutta-700 030, India\Arefs{n}}
\item \Idef{dubna}{Joint Institute for Nuclear Research, 141980 Dubna, Moscow region, Russia\Arefs{o}}
\item \Idef{erlangen}{Universit\"at Erlangen--N\"urnberg, Physikalisches Institut, 91054 Erlangen, Germany\Arefs{l}}
\item \Idef{freiburg}{Universit\"at Freiburg, Physikalisches Institut, 79104 Freiburg, Germany\Arefs{l}\Arefs{s}}
\item \Idef{cern}{CERN, 1211 Geneva 23, Switzerland}
\item \Idef{liberec}{Technical University in Liberec, 46117 Liberec, Czech Republic\Arefs{m}}
\item \Idef{lisbon}{LIP, 1000-149 Lisbon, Portugal\Arefs{p}}
\item \Idef{mainz}{Universit\"at Mainz, Institut f\"ur Kernphysik, 55099 Mainz, Germany\Arefs{l}}
\item \Idef{miyazaki}{University of Miyazaki, Miyazaki 889-2192, Japan\Arefs{q}}
\item \Idef{moscowlpi}{Lebedev Physical Institute, 119991 Moscow, Russia}
\item \Idef{munichtu}{Technische Universit\"at M\"unchen, Physik Dept., 85748 Garching, Germany\Arefs{l}\Arefs{r}}
\item \Idef{nagoya}{Nagoya University, 464 Nagoya, Japan\Arefs{q}}
\item \Idef{praguecu}{Charles University in Prague, Faculty of Mathematics and Physics, 18000 Prague, Czech Republic\Arefs{m}}
\item \Idef{praguectu}{Czech Technical University in Prague, 16636 Prague, Czech Republic\Arefs{m}}
\item \Idef{protvino}{State Scientific Center Institute for High Energy Physics of National Research Center `Kurchatov Institute', 142281 Protvino, Russia}
\item \Idef{saclay}{IRFU, CEA, Universit\'e Paris-Saclay, 91191 Gif-sur-Yvette, France\Arefs{s}}
\item \Idef{taipei}{Academia Sinica, Institute of Physics, Taipei 11529, Taiwan\Aref{tw}}
\item \Idef{telaviv}{Tel Aviv University, School of Physics and Astronomy, 69978 Tel Aviv, Israel\Arefs{t}}
\item \Idef{triest_u}{University of Trieste, Dept.\ of Physics, 34127 Trieste, Italy}
\item \Idef{triest_i}{Trieste Section of INFN, 34127 Trieste, Italy}
\item \Idef{triest_ictp}{Abdus Salam ICTP, 34151 Trieste, Italy}
\item \Idef{turin_u}{University of Turin, Dept.\ of Physics, 10125 Turin, Italy}
\item \Idef{turin_i}{Torino Section of INFN, 10125 Turin, Italy}
\item \Idef{illinois}{University of Illinois at Urbana-Champaign, Dept.\ of Physics, Urbana, IL 61801-3080, USA\Aref{nsf}}
\item \Idef{warsaw}{National Centre for Nuclear Research, 00-681 Warsaw, Poland\Arefs{u}}
\item \Idef{warsawu}{University of Warsaw, Faculty of Physics, 02-093 Warsaw, Poland\Arefs{u}}
\item \Idef{warsawtu}{Warsaw University of Technology, Institute of Radioelectronics, 00-665 Warsaw, Poland\Arefs{u} }
\item \Idef{yamagata}{Yamagata University, Yamagata 992-8510, Japan\Arefs{q} }
\end{Authlist}
%
%
\renewcommand\theenumi{\alph{enumi}}
\begin{Authlist}
\item [{\makebox[2mm][l]{\textsuperscript{\#}}}] Corresponding authors
\item [{\makebox[2mm][l]{\textsuperscript{*}}}] Deceased
\item \Adef{a}{Also at Instituto Superior T\'ecnico, Universidade de Lisboa, Lisbon, Portugal}
\item \Adef{b}{Also at Dept.\ of Physics, Pusan National University, Busan 609-735, Republic of Korea and at Physics Dept., Brookhaven National Laboratory, Upton, NY 11973, USA}
\item \Adef{r}{Supported by the DFG cluster of excellence `Origin and Structure of the Universe' (www.universe-cluster.de) (Germany)}
\item \Adef{p2i}{Supported by the Laboratoire d'excellence P2IO (France)}
\item \Adef{d}{Also at Chubu University, Kasugai, Aichi 487-8501, Japan\Arefs{q}}
\item \Adef{x}{Also at Dept.\ of Physics, National Central University, 300 Jhongda Road, Jhongli 32001, Taiwan}
\item \Adef{e}{Also at KEK, 1-1 Oho, Tsukuba, Ibaraki 305-0801, Japan}
\item \Adef{g}{Also at Moscow Institute of Physics and Technology, Moscow Region, 141700, Russia}
\item \Adef{v}{Supported by Presidential Grant NSh--999.2014.2 (Russia)}
\item \Adef{h}{Present address: RWTH Aachen University, III.\ Physikalisches Institut, 52056 Aachen, Germany}
\item \Adef{y}{Also at Dept.\ of Physics, National Kaohsiung Normal University, Kaohsiung County 824, Taiwan}
\item \Adef{i}{Present address: Uppsala University, Box 516, 75120 Uppsala, Sweden}
\item \Adef{c}{  Supported by the DFG Research Training Group Programmes 1102 and 2044 (Germany)} 
%
%
\item \Adef{l}{  Supported by BMBF - Bundesministerium f\"ur Bildung und Forschung (Germany)}
\item \Adef{s}{  Supported by FP7, HadronPhysics3, Grant 283286 (European Union)}
\item \Adef{m}{  Supported by MEYS, Grant LG13031 (Czech Republic)}
\item \Adef{n}{  Supported by SAIL (CSR) and B.Sen fund (India)}
\item \Adef{o}{  Supported by CERN-RFBR Grant 12-02-91500}
\item \Adef{p}{\raggedright 
                 Supported by FCT - Funda\c{c}\~{a}o para a Ci\^{e}ncia e Tecnologia, COMPETE and QREN, Grants CERN/FP 116376/2010, 123600/2011 
                 and CERN/FIS-NUC/0017/2015 (Portugal)}
\item \Adef{q}{  Supported by MEXT and JSPS, Grants 18002006, 20540299 and 18540281, the Daiko and Yamada Foundations (Japan)}
\item \Adef{tw}{ Supported by the Ministry of Science and Technology (Taiwan)}
\item \Adef{t}{  Supported by the Israel Academy of Sciences and Humanities (Israel)}
\item \Adef{nsf}{Supported by NSF - National Science Foundation (USA)}
\item \Adef{u}{  Supported by NCN, Grant 2015/18/M/ST2/00550 (Poland)}
\end{Authlist}
%
%
%

\end{titlepage}

\section{ Introduction}
\label{chapter:introduction}
An interesting and recently examined property of the quark distribution in a nucleon that is polarised transversely to its momentum is the fact that it is not left-right symmetric with respect to the plane defined by the directions of nucleon spin and momentum. This asymmetry of the distribution function is called the Sivers effect and  was first suggested~\cite{Sivers:1989cc} as an explanation for the large left-right single transverse spin asymmetries observed for pions produced in the reaction $p^{\uparrow} p \rightarrow \pi X$~\cite{Antille:1980th,Adams:1991cs,Adams:1991rw}.  On the basis of T-invariance arguments the existence of such an asymmetric distribution, known as Sivers distribution function, was originally excluded~\cite{Collins:1992kk}.  Ten years later it was recognised however that it was indeed possible~\cite{Brodsky:2002cx}. At that time it was also predicted that
the Sivers function in semi-inclusive measurements of hadron production in DIS (SIDIS) and in the Drell-Yan
process have opposite sign~\cite{Collins:2002kn}, a property referred to as ``restricted universality''.
A few years later the Sivers effect was experimentally observed in SIDIS experiments on transversely polarised proton targets, first by the HERMES Collaboration~\cite{Airapetian:2004tw} and then
by the COMPASS Collaboration~\cite{Alekseev:2010rw}. Using the first HERMES data and the early COMPASS data taken with a transversely polarised deuteron target~\cite{Alexakhin:2005iw}, a combined analysis soon allowed for first extractions of the Sivers function for u and d-quarks~\cite{Vogelsang:2005cs, Efremov:2008vf, Anselmino:2008sga}. More precise measurements of the Sivers effect were performed since by the HERMES~\cite{Airapetian:2009ae} and COMPASS~\cite{Ageev:2006da,Alekseev:2008aa, Adolph:2012sp} Collaborations, and new measurements with a transversely polarised $^3$He target were also carried out at JLab~\cite{Qian:2011py, Zhao:2014qvx}. More information can be found in recent reviews~\cite{Barone:2010zz,Aidala:2012mv,Avakian:2016rst}.\par
At this point the question arises whether the gluon distribution in a transversely polarised nucleon is left-right symmetric or exhibits a Sivers effect similar to the quark distributions. Recently, the issue has been discussed repeatedly in the literature and the properties of the gluon Sivers distributions have been studied in great detail~\cite{Mulders:2000sh,Boer:2015vso}. While it was found that a non-zero Sivers function implies motion of partons in the nucleon, presently the connection between the Sivers function and the parton orbital angular momentum in the nucleon can only be described in a model-dependent way~\cite{Burkardt:2002ks}. The correspondence between the Sivers effect and the transverse motion of partons
has been originally proposed by M.~Burkardt~\cite{Burkardt:2002hr, Burkardt:2003uw, Burkardt:2003je}. Hence it is of great interest to know whether there exists a gluon Sivers effect or not.\par 
Presently, little is known on the gluon Sivers function. An important theoretical constraint comes from the so-called Burkardt sum rule~\cite{Burkardt:2004ur}. It states, based on the presence of QCD colour-gauge links, that the total transverse momentum of all partons inside a transversely polarised proton should vanish. Fits to the Sivers asymmetry using SIDIS data~\cite{Anselmino:2008sga} almost fulfil, within uncertainties, the Burkardt sum rule, leaving little space for a gluon contribution.
From the null result of the COMPASS experiment for the Sivers asymmetry of positive  and negative hadrons produced on a transversely polarised deuteron target~\cite{Alexakhin:2005iw}, together with additional theoretical considerations, Brodsky and Gardner~\cite{Brodsky:2006ha} stated that the gluon contribution to the parton orbital angular momentum  should be negligible, and consequently that the gluon Sivers effect should be small. Also, using the so-called transverse momentum dependent (TMD) generalised parton model and the most recent phenomenological information on the quark Sivers distributions coming from SIDIS data, interesting constraints on our knowledge of the gluon Sivers function were derived~\cite{D'Alesio:2015uta} from the recent precise data on the transverse single spin asymmetry $A_N(p^{\uparrow} p\rightarrow \pi^0X)$ measured at central rapidity by the PHENIX Collaboration at RHIC~\cite{Adare:2013ekj}.\par 
In  DIS,  the  leading-order  virtual-photon  absorption  process  (LP)  does  not provide  direct  access  to  the gluon  distribution  since  the  virtual-photon  does  not couple  to the  gluon, so that higher-order processes have to be studied, {\it i.e.} QCD Compton scattering (QCDC) and Photon-Gluon Fusion (PGF). It is well known that in lepton-proton scattering one of the most promising processes to directly probe the gluon is open charm production, $\ell p^{\uparrow} \rightarrow \ell' c\bar{c} X$.  This is the channel that the COMPASS Collaboration has investigated at length in order to measure $\Delta g/g$ , the gluon polarisation in a longitudinally polarised nucleon~\cite{Adolph:2012ca}.  Tagging the charm quark by  identifying D-mesons in the final state has the advantage that in the lowest order of the strong coupling constant there are no other contributions to the cross section and one becomes essentially sensitive to the gluon distribution function. An alternative method to tag the gluon in DIS, which has the advantage of higher statistics, has also been developed and used by COMPASS, {\it i.e.} the production of high-$p_T$ hadrons~\cite{Bravar:1997kb, Adolph:2012vj}. In the LP,  the  hadron  transverse  momentum $p_T$ with respect  to  the  virtual photon direction (in the frame where the nucleon momentum is parallel to this direction) originates from the intrinsic transverse momentum $k_T$ of the struck quark in the nucleon  and its fragmentation, which both lead to a small transverse component. On the contrary, both the QCDC and PGF hard processes can provide hadrons with high transverse momentum. Therefore, tagging events with hadrons of high transverse momentum $p_T$ enhances the contribution of higher-order processes. Nevertheless, although in the high-$p_T$ sample the PGF fraction is enriched, in order to single out the contribution of the PGF process to the measured asymmetry the contributions from LP and QCDC have to be subtracted~\cite{Adolph:2015cvj}.\par
In this letter, the gluon Sivers effect is investigated using COMPASS data collected by scattering a 160~GeV/$c$ muon beam off transversely polarised deuterons and protons. 
The experimental set-up and the data selection are described in Section~2. In Section~3 the measurement is described. The details of the analysis are given in Section~4. The procedure of neural network  (NN) training with a Monte Carlo data sample is shown in Section~5. Section~6 contains the overview of the systematic studies. In Section~7 the results are presented. Summary and conclusions are given in Section~8.

\section{Experimental set-up and data samples}
\label{sect:setup}
The COMPASS experiment uses a fixed target set-up and a polarised muon beam delivered by the M2 beam line of the CERN SPS. The transversely polarised deuteron target used for the 2003 and 2004 data taking consisted of two oppositely polarised cylindrical cells situated along the beam, each 60~cm long with a 10~cm gap in between. In 2010 the transversely polarised proton target consisted of three cells: 30~cm, 60~cm and 30~cm long with the central cell oppositely polarised to the downstream and upstream cell and 5~cm gaps between the cells. During all data taking periods the polarisation was reversed once per week, in this way systematic effects due to acceptance are cancelled. For the deuteron runs the target was filled with $^6$LiD. The $^6$Li nucleus can be regarded as one quasi-free deuteron and a $^4$He core. The average dilution factor $f_d$, defined as the ratio of the DIS cross section on polarisable nucleons in the target to the cross section on all target nucleons, amounts to $0.36$ and includes also electromagnetic radiative corrections. The average polarisation of the deuteron was $0.50$. For the asymmetry measurements on the proton, NH$_3$ was used as a target. Its average dilution factor $f_p$ amounts to $0.15$ and the proton polarisation to $0.80$. In both cases, the naturally polarised muon beam of 160 GeV/$c$ was used. The basic features of the COMPASS spectrometer, as described in Ref.~\cite{Abbon:2007pq}, are the same for 2003-4 and 2010 data taking. Several upgrades were performed in 2005, the main one being the installation of a  new target magnet, which allowed to increase the polar angle acceptance from 70~mrad to 180~mrad.\par
A crucial point of this analysis is the search for an observable that is strongly correlated with the gluon azimuthal angle $\phi_g$. In the LEPTO generator~\cite{Ingelman:1996mq}, gluons are accessed via PGF with a quark-antiquark pair in the final state and the fragmentation process is described by the Lund model~\cite{Andersson:1983ia}. 
As a result of MC studies, the best correlation is found between $\phi_g$ and $\phi_P$, where the latter denotes the azimuthal angle of the
vector sum $\boldsymbol{P}$ of the two hadron momenta. For the present analysis, two charged hadrons for each event are selected. If more than two charged hadrons are reconstructed in an event, only the hadron with the largest transverse momentum, $p_{T1}$, and the one with the second-largest transverse momentum, $p_{T2}$, are taken into account. In order to enhance the PGF
fraction in the sample and at the same time the correlation between 
$\phi_g$ and $\phi_P$, a further requirement is applied to the transverse momenta of the two hadrons:
$p_{T1} > 0.7$~GeV/$c$ and $p_{T2} > 0.4$~GeV/$c$. Moreover, the fractional energies of the two hadrons must fulfil the following conditions: $z_i>0.1$ $(i=1,2)$ and $z_1+z_2<0.9$, where the last requirement rejects events from diffractive vector meson production. Hadron pairs are selected with no charge constraint. With this choice the correlation coefficient is 0.54.
The Sivers asymmetry is then obtained as the sine modulation in the Sivers angle, $\phi_{Xiv} =
\phi_P - \phi_S$. Here $\phi_S$ is the azimuthal angle of the nucleon spin vector.\par 
The same kinematic data selection is used for both deuteron and  proton data. The requirement on photon virtuality, ~{${Q^2>1}~($GeV/$c)^2$}, selects events in the perturbative region and the one on the mass of the hadronic final state, ~{$W>5~$GeV/$c^2$}, removes the region of the exclusive nucleon resonance production. The Bjorken-$x$ variable covers the range ~${0.003<x_{Bj}<0.7}$. For the fractional energy of the virtual photon, $y$, the limit ~$y>0.1$ removes a region sensitive to experimental biases and the requirement $y<0.9$ rejects events with large electromagnetic radiative corrections.

\section{Sivers asymmetry in two hadron production}
\label{sect:3processes}
In order to extract the gluon Sivers asymmetry, $\mu+N \rightarrow \mu' + 2h + X$ events are selected as described in Section \ref{sect:setup}. By labelling with the symbol  $\uparrow$  the cross section associated to a target cell polarised upwards in the laboratory and by $\downarrow$ the cross section of a target cell polarised downwards in the laboratory, the Sivers asymmetry can be written as  
\begin{equation}
    A_T^{2h}(\vec{x},\phi_{Siv})=\frac{\Delta\sigma(\vec{x},\phi_{Siv})}{\sigma(\vec{x})},
\label {eq:asymmetry}
\end{equation}
where $\vec{x} = (x_{Bj}, Q^2, p_{T1}, p_{T2}, z_1, z_2)$, $\Delta\sigma\equiv d^7\sigma{\uparrow}-d^7\sigma{\downarrow}$ and $\sigma\equiv d^7\sigma{\uparrow}+d^7\sigma{\downarrow}$. All cross sections are integrated over the two azimuthal angles $\phi_S$ and $\phi_R$, where $\phi_R$ is the azimuthal angle of the relative momentum of the two hadrons, $\boldsymbol{R}=\boldsymbol{P}_1-\boldsymbol{P}_2$. The number of events in a $\phi_{Siv}$ bin is given by
\begin{equation}
    N(\vec{x}, \phi_{Siv}) = \alpha(\vec{x}, \phi_{Siv})\left(1+fP_TA^{Siv}(\vec{x})\sin{\phi_{Siv}}\right).
\label{eq:modulation}
\end{equation} 
Here $f$ is the dilution factor, $P_T$ the target polarisation and $\alpha = an\Phi\sigma_0$ an acceptance-dependent factor, where $a$ is the total spectrometer acceptance, $n$ the density of scattering centres, $\Phi$ the beam flux and $\sigma_0$ the spin-averaged part of the cross section.  From here on, the Sivers asymmetry $A_T^{2h}(\vec{x}, \phi_{Siv})$ is factorised into the azimuth-independent amplitude $A^{Siv}(\vec{x})$ and the modulation $\sin{\phi_{Siv}}$.\\
In order to extract the Sivers asymmetry of the gluon, the amplitude of the $\sin{\phi_{Siv}}$ modulation is extracted from data. The general expression for the cross section of SIDIS production with at least one hadron in the final state is well
known~\cite{Bacchetta:2006tn}. It contains eight azimuthal
modulations, which are functions of the single-hadron azimuthal angle and $\phi_S$. In the absence of correlations possibly introduced by experimental effects, they are all orthogonal, so that the
Sivers asymmetry can either be extracted as the amplitude of the $\sin{\phi_{Siv}}$ modulation
or one can perform a simultaneous fit of all  eight amplitudes. For the case of heavy-quark pair and dijet production in lepton-nucleon collisions, all azimuthal asymmetries associated to the gluon distribution function have been recently worked out in Ref. \citenum{Boer:2016fqd}. There, the  Sivers asymmetry is defined as the amplitude of the $\sin{(\phi_T-\phi_S)}$ modulation, where $\phi_T$ is the azimuthal angle of the transverse-momentum vector of the quark-antiquark pair, $\boldsymbol{q}_T$. In our analysis, $\phi_T$ is  replaced by $\phi_P$, due to its correlation with the gluon azimuthal angle $\phi_g$, and the Sivers asymmetry is extracted taking into account only the $\sin{(\phi_P-\phi_S)}$ modulation in the cross section. It has been verified that including in the cross section  the same eight transverse-spin modulations as
in SIDIS single-hadron production~\cite{Bacchetta:2006tn} and extracting simultaneously all asymmetries gives the same result on the gluon Sivers asymmetry.\par
\begin{figure}[tbp]
    \centering
    \begin{subfigure}[b!]{0.25\textwidth}
    \centering
    \includegraphics[width=\textwidth]{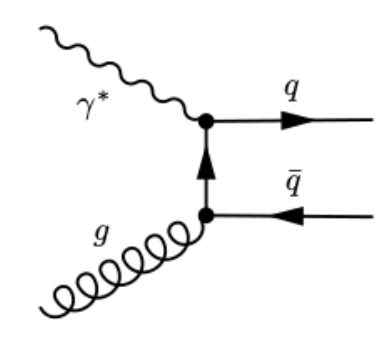}
    \caption{ }
    \end{subfigure}
    \begin{subfigure}[b!]{0.25\textwidth}
    \centering
    \includegraphics[width=\textwidth]{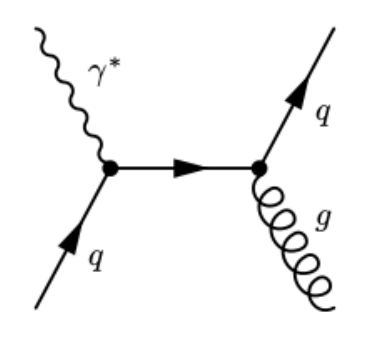}
    \caption{ }
    \end{subfigure}
    \begin{subfigure}[b!]{0.25\textwidth}
    \centering
    \includegraphics[width=\textwidth]{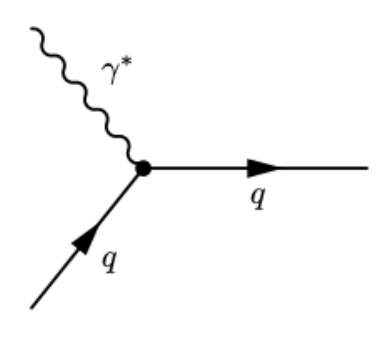}
    \caption{ }
    \end{subfigure}
    \caption{Feynman diagrams considered for $\gamma^*N$ scattering: a) photon-gluon fusion (PGF), b) gluon radiation (QCD Compton scattering), c) Leading order process (LP).}\label{fig:processes}
\end{figure}

  
In order to determine the Sivers asymmetry for gluons from two-hadron production in SIDIS, it is necessary to assume that the main contributors to muon-nucleon DIS are the three processes (Fig.~\ref{fig:processes}) as presented in Ref.~\cite{Ingelman:1996mq}.
This model is successful in describing the unpolarised data. At COMPASS kinematics, the leading process appears at zero-order QCD in the total DIS cross section and it is the dominant process, while the other two processes, photon-gluon fusion and QCD Compton, are first-order QCD processes and hence suppressed. However, their contribution can be enhanced by constraining the transverse momentum of the produced hadrons, as mentioned above. \par
Introducing the process fractions $R_j=\sigma_j/\sigma$ ($j\in$ \{PGF, QCDC, LP\}), the amplitude of the Sivers asymmetry can be expressed in terms of the amplitudes of the three contributing processes:
\begin {equation}
\begin {split}
    fP_TA^{Siv}\sin{\phi_{Siv}} &=\frac{\Delta\sigma}{\sigma}=\frac{\sigma_{\text{PGF}}}{\sigma}\frac{\Delta\sigma_{\text{PGF}}}{\sigma_{\text{PGF}}}+\frac{\sigma_{\text{QCDC}}}{\sigma}\frac{\Delta\sigma_{\text{QCDC}}}{\sigma_{\text{QCDC}}}+\frac{\sigma_{\text{LP}}}{\sigma}\frac{\Delta\sigma_{\text{LP}}}{\sigma_{\text{LP}}}\\&=fP_T (R_{\text{PGF}}A_{\text{PGF}}^{Siv}+R_{\text{QCDC}}A_{\text{QCDC}}^{Siv}+R_{\text{LP}}A_{\text{LP}}^{Siv}) \sin{\phi_{Siv}},
\end{split}
    \label{eq:decomposition}
\end {equation}
with $\sigma=\sum_j\sigma_j$, $\Delta\sigma=\sum_j\Delta\sigma_j$ and $fP_TA^{Siv}_j\sin{\phi_{Siv}}=\Delta\sigma_j/\sigma_j$. The determination of $R_j$ is done on an event-by-event basis by using the neural networks (NN) trained on Monte Carlo data as described in Section~5.

\section{Asymmetry extraction using the methods of weights}
\label{sect:weighted_method}
The method adopted in the present analysis was already applied to extract the gluon polarisation from the longitudinal 
double-spin asymmetry in the SIDIS measurement of single-hadron production~\cite{Adolph:2015cvj}. 
Both for the deuteron data (two target cells) and the proton data (three target cells), 
four target configurations can be introduced. 
In the case of the two-cell target: 1 - upstream, 2 - downstream, 3 - upstream$'$, 4 - downstream$'$. 
In the case of the three-cell target: 1 - (upstream+downstream), 2 - centre, 3 - (upstream$'$+downstream$'$), 4 - centre$'$. 
Here upstream$'$, centre$'$ and downstream$'$ denote the cells after the polarisation reversal and 
configuration 1 has the polarisation pointing upwards in the laboratory frame. Decomposing the Sivers asymmetry 
into the asymmetries of the contributing processes (Eq.~\eqref{eq:decomposition})  
and introducing the Sivers modulation $\beta^t_j(\vec{x}, \phi_{\text{Siv}})=R_j(\vec{x})f(\vec{x})P^t_T\sin{ \phi_{\text{Siv}}}$, which is specific for process $j$, one can rewrite Eq.~\eqref{eq:modulation}:
\begin{equation}
\begin{split}
   N^t(\vec{x},  \phi_{\text{Siv}})=\alpha^t(\vec{x},  \phi_{\text{Siv}})\Big(&1+\beta^t_{\text{PGF}}(\vec{x}, \phi_{\text{Siv}})A^{Siv}_{\text{PGF}}(\vec{x})\\&+\beta^t_{\text{QCDC}}(\vec{x}, \phi_{\text{Siv}})A^{Siv}_{\text{QCDC}}(\vec{x})+\beta^t_{\text{LP}}(\vec{x}, \phi_{\text{Siv}})A^{Siv}_{\text{LP}}(\vec{x})\Big),
\end{split}
\label{eq:events_number}
\end{equation}
where $t = 1, 2, 3, 4$ denotes the target configuration. 

In order to minimise statistical uncertainties for each process, a weighting factor is introduced. 
It is known~\cite{Pretz:2011qb} that the choice  $\omega_j=\beta_j$ for the weight optimises the statistical uncertainty
but variations of the target polarisation $P_T$ in time may introduce a bias to the final result.
Therefore, the weighting factor $\omega_j\equiv\beta_j/P_T$ is used instead.  
Each of the four equations \eqref{eq:events_number} is weighted three times with $\omega_j$ 
depending on the process $j\in\{\text{PGF, QCDC, LP}\}$ and integrated over $ \phi_{\text{Siv}}$ and $\vec{x}$, yielding twelve observed quantities $q_j^t$:

    \begin{equation}
        \begin{split}
           q_j^t&=\int d\vec{x}d \phi_{\text{Siv}}\omega_j(\vec{x}, \phi_{\text{Siv}})N^t(\vec{x},  \phi_{\text{Siv}}) \\
           &=\tilde{\alpha}_j^t\Big(1+\{\beta^t_{\text{PGF}}\}_{\omega_j}\big\{A_{\text{PGF}}^{Siv}\big\}_{\beta_{\text{PGF}}\omega_j} +\{\beta^t_{\text{QCDC}}\}_{\omega_j}\big\{A_{\text{QCDC}}^{Siv}\big\}_{\beta_{\text{QCDC}}\omega_j}+\{\beta^t_{\text{LP}}\}_{\omega_j}\big\{A_{\text{LP}}^{Siv}\big\}_{\beta_{\text{LP}}\omega_j}\Big),\\ 
        \label{eq:weighted_processes}
        \end{split} 
     \end{equation}
where $\tilde{\alpha}_j^t$ is the $\omega_j$-weighted acceptance-dependent factor.
The quantities $\{\beta_i^t\}_{\omega_j}$ and $\{A_i^{\text{Siv}} \}_{\beta^t_i \omega_j}$ are weighted
averages, where the weight factor is denoted in the subscript.%
\footnote{$\{\beta\}_{\omega}=\frac{\int\alpha\beta\omega d\vec{x}d
\phi_{\text{Siv}}}{\int\alpha\omega d\vec{x}d
\phi_{\text{Siv}}},~\{A\}_{\beta\omega}=\frac{\int A\alpha\beta\omega d\vec{x}d
\phi_{\text{Siv}}}{\int\alpha\beta\omega d\vec{x}d \phi_{\text{Siv}}}$}

The acceptance factors $\tilde{\alpha}_j^t$ cancel when for asymmetry extraction one uses the double ratio 
\begin{equation}
r_j:=\frac{q^1_j q^4_j}{q^2_j q^3_j}
\label{eq:DR}
\end{equation} 
as the data taking was performed such that
$\tilde{\alpha}_j^1 \tilde{\alpha}_j^4 /  \tilde{\alpha}_j^2 \tilde{\alpha}_j^3=1$.
If this condition is not fulfilled, false asymmetries may occur. 
It is checked that this is not the case (see Section \ref{sect:systematics}). \\

In the analysis, the quantities $q_j$ and $\{\beta^t_{i}\}_{\omega_j}$ are approximated as follows:

\begin {equation}
q_j^t\approx\sum_{k=1}^{N_t}\omega_j^k,
\end{equation}
\begin{equation}
\{\beta_i^t\}_{\omega_j} \approx\frac{\sum\limits^{N^t}_{k=1} \beta_i^{t,k} \omega^k_j}{\sum\limits^{N^t}_{k=1}\omega^k_j}.
\label{eq:mean}
\end{equation}
The latter approximation holds for small observed raw asymmetries, {\it i.e.} $\omega A\ll 1$.
In order to avoid numerical
inconsistencies in Eq.~\eqref{eq:mean} due to a zero-pole when integrating over the full range of $\phi_{\text{Siv}}$,\footnote{ 
Note that $\omega^k_j$, which contains $\sin{ \phi_{\text{Siv}}}$,  is integrated in the region 0 to $2 \pi$. } two bins in $\phi_{\text{Siv}}$ $([0; \pi],[\pi; 2\pi])$ are introduced.
In the aforementioned three double ratios given in Eq.~(\ref{eq:DR})
only asymmetries are unknown.
However, in order to solve the system of equations one needs to assume that the
weighted asymmetry for a given process $i$ is the same for the three different weights
$\omega_{j}\beta_i$, {\it i.e.}
$\{A_{i}\}_{\beta_{i} \omega_{\text{PGF}}}=\{A_{i}\}_{\beta_{i} \omega_{\text{QCDC}}}=\{A_{i}\}_{\beta_{i} \omega_{\text{LP}}}$. 
This means that the values of  $\omega_{j}$ and $A_{i}$ must be uncorrelated. 
For example, since $\omega_{j}$ is proportional to $R_j$, which strongly depends on the hadron transverse momentum,
one has to use a kinematic region where the asymmetries $A_{i}$ are expected to be independent of $p_T$.
Under these assumptions, the number of unknown weighted asymmetries is three, 
which exactly corresponds to the number of equations of type (\ref{eq:DR}). 
These equations are solved by a $\chi^2$ fit that includes simultaneously both bins in  $ \phi_{\text{Siv}}$.

Assuming that $A_i$ can be approximated 
by a linear function of $x_i$
and that $x_i$ is not correlated with $\omega_{j}$, results in
\begin{equation}
\{A_{i}\}_{\beta_{i} \omega_{i} } =A_{i}(\{x_{i}\}_{\beta_{i} \omega_{i}}).
\end{equation}
This approximation allows to interpret the obtained results as an asymmetry value
measured at the weighted value of $x_i$. For each process, the weighted
value of $x_i$ is obtained from MC using the relation
\begin{equation}
\{x_{i}\}_{\beta_{i} \omega_{i}} = \frac{ \sum\limits_{k=1}^{N_i} x_i^k \beta_{i}^{k} \omega_{i}^k }{ \sum\limits_{k=1}^{N_i} 
\beta_{i}^k \omega_{i}^{k}}.
\end{equation} 
Here, $N_i$ is the number of events of type $i$ in MC data. 
The assumption that the values of $x_i$ are not correlated with $\omega_{j}$,
which allows us to consider only $\{x_{i}\}_{\omega_{i}\beta_{i}}$,
was verified using MC data.  
The details of the analysis are given in~\cite{SzabelskiPHD}.

\section{Monte Carlo optimisation and Neural Network training}
\label{sect:NN_training}
The present analysis is very similar to the one used for the $\Delta g/g$ extraction from high-$p_T$ hadron pairs~\cite{Adolph:2012vj} and single hadrons~\cite{Adolph:2015cvj}. For the NN training with custom input, output and target vector the package NetMaker~\cite{NetMaker} is used. The NN is trained with a Monte Carlo sample with process identification. As input vector the following six kinematic variables are chosen: $x_{Bj}, Q^2, p_{T1}, p_{T2}, p_{L1}, p_{L2}$. The latter two are the longitudinal components of the hadron momenta. The trained neural network is applied to the data by taking the vector of the aforementioned six variables,  and its output is interpreted as probabilities that the given event is a result of one of the three contributing processes. Hence the simulated distributions of these variables need to be in agreement with the corresponding distributions in the data samples.\par
\begin{figure}[tbp]
		\begin{tabular}{ccc}			
			\includegraphics[width=0.32\textwidth]{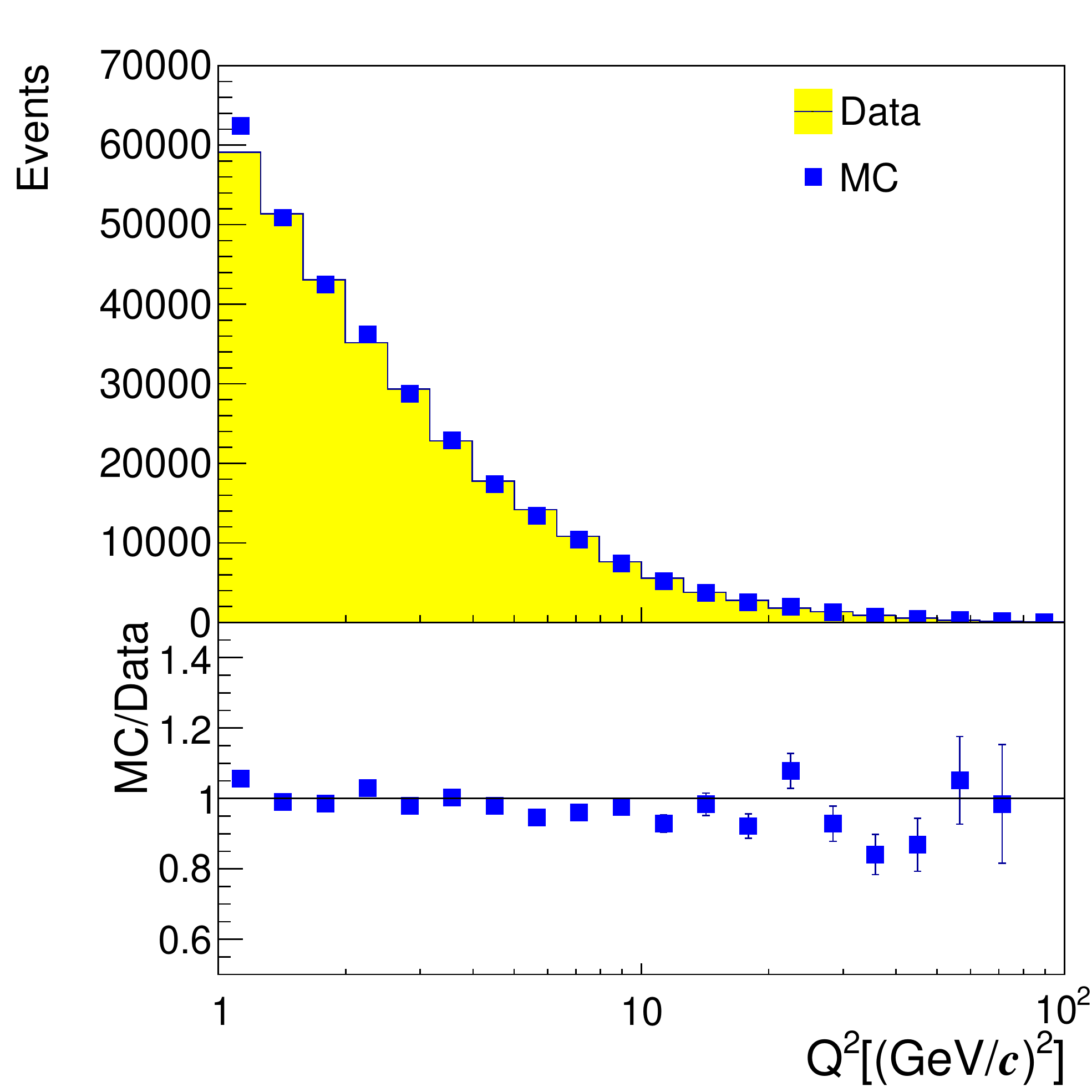}%
			&\includegraphics[width=0.32\textwidth]{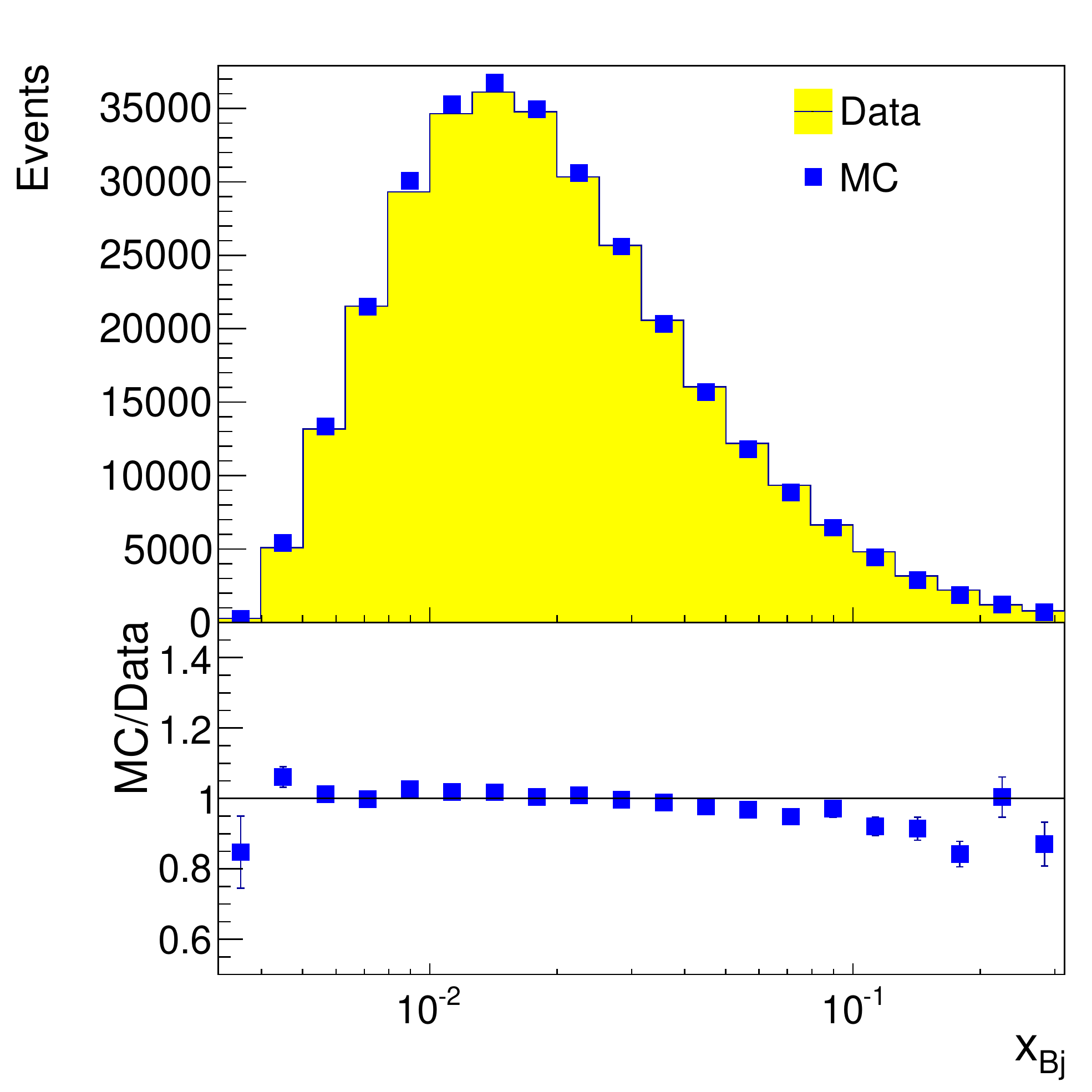}%
			&\includegraphics[width=0.32\textwidth]{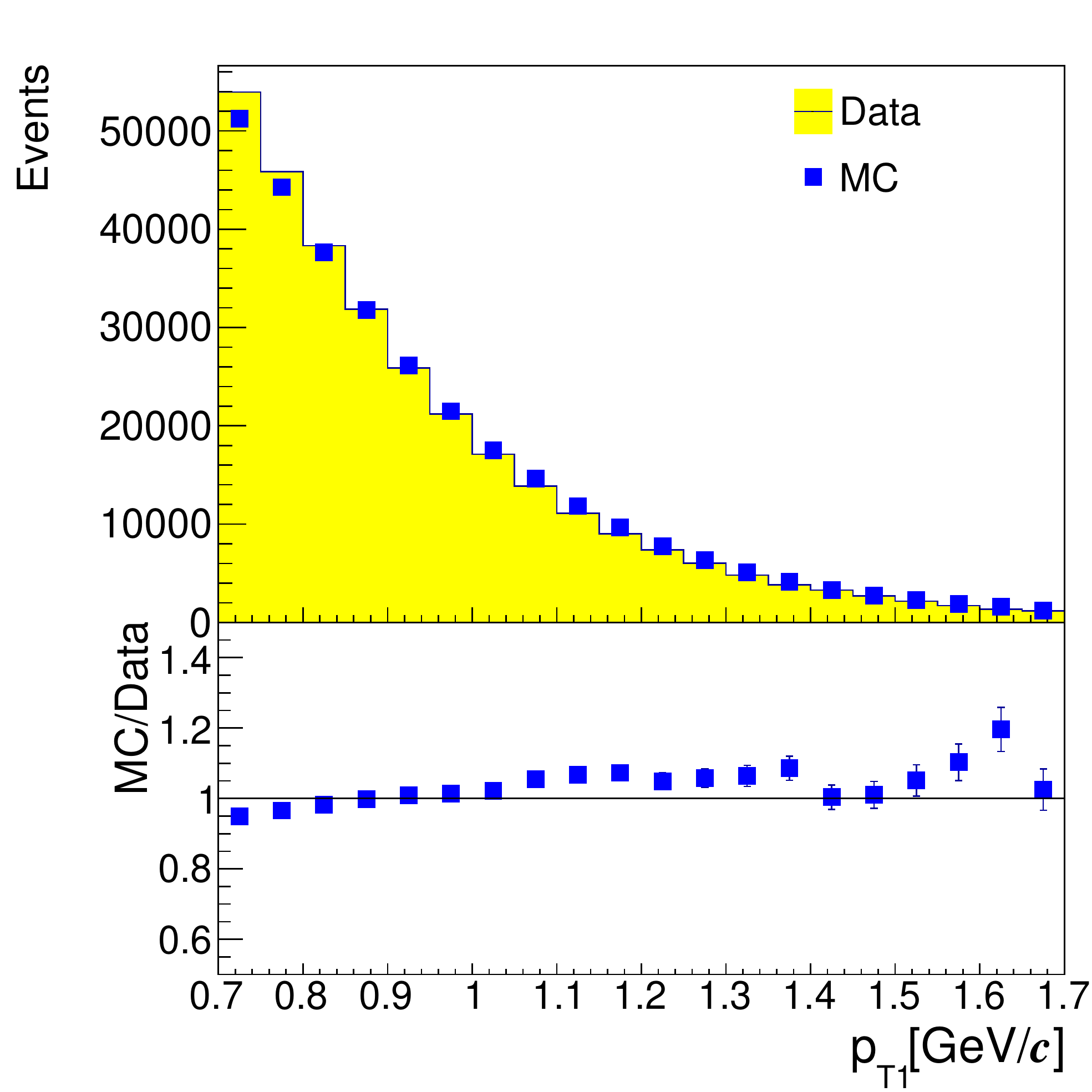}\\
			\includegraphics[width=0.32\textwidth]{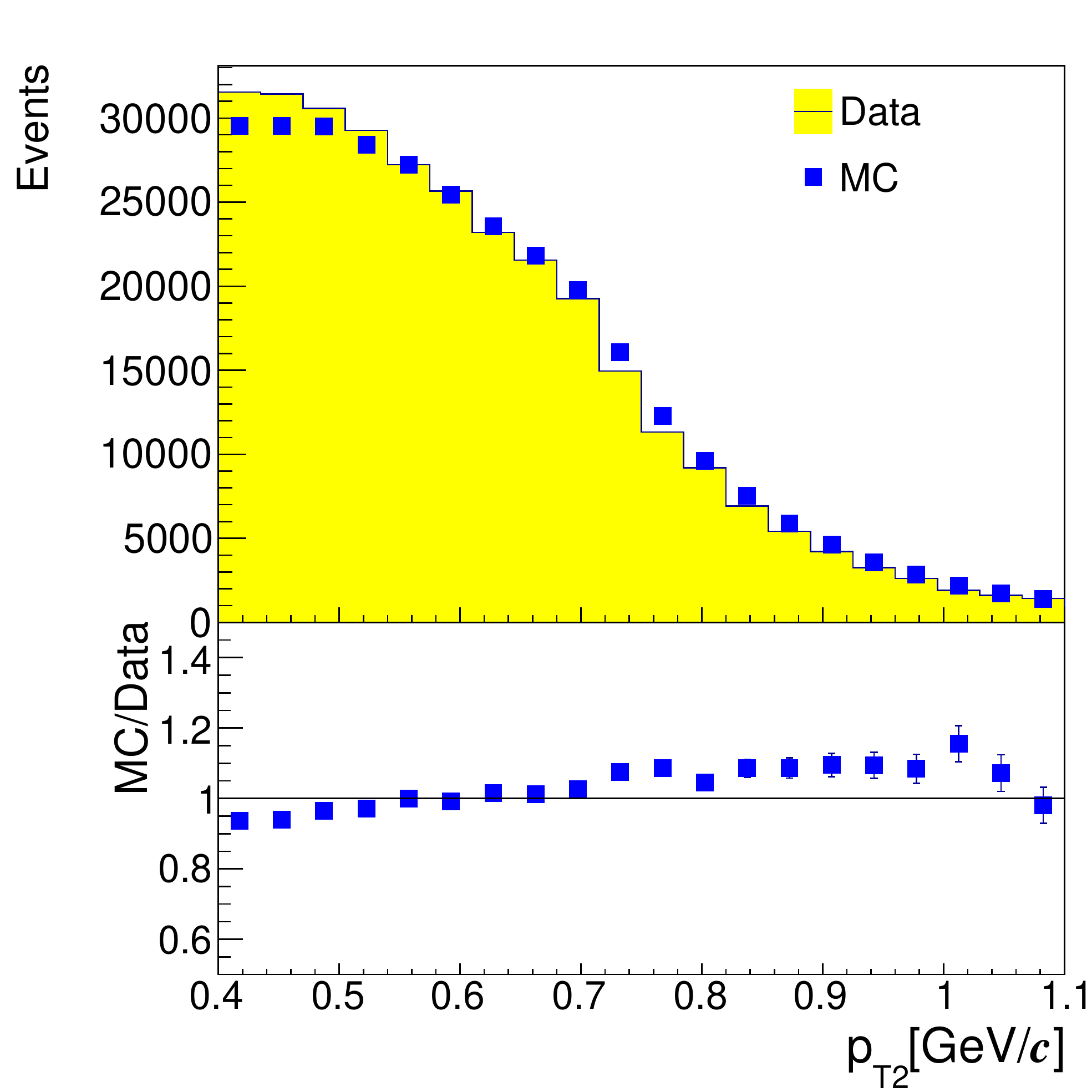}%
            &\includegraphics[width=0.32\textwidth]{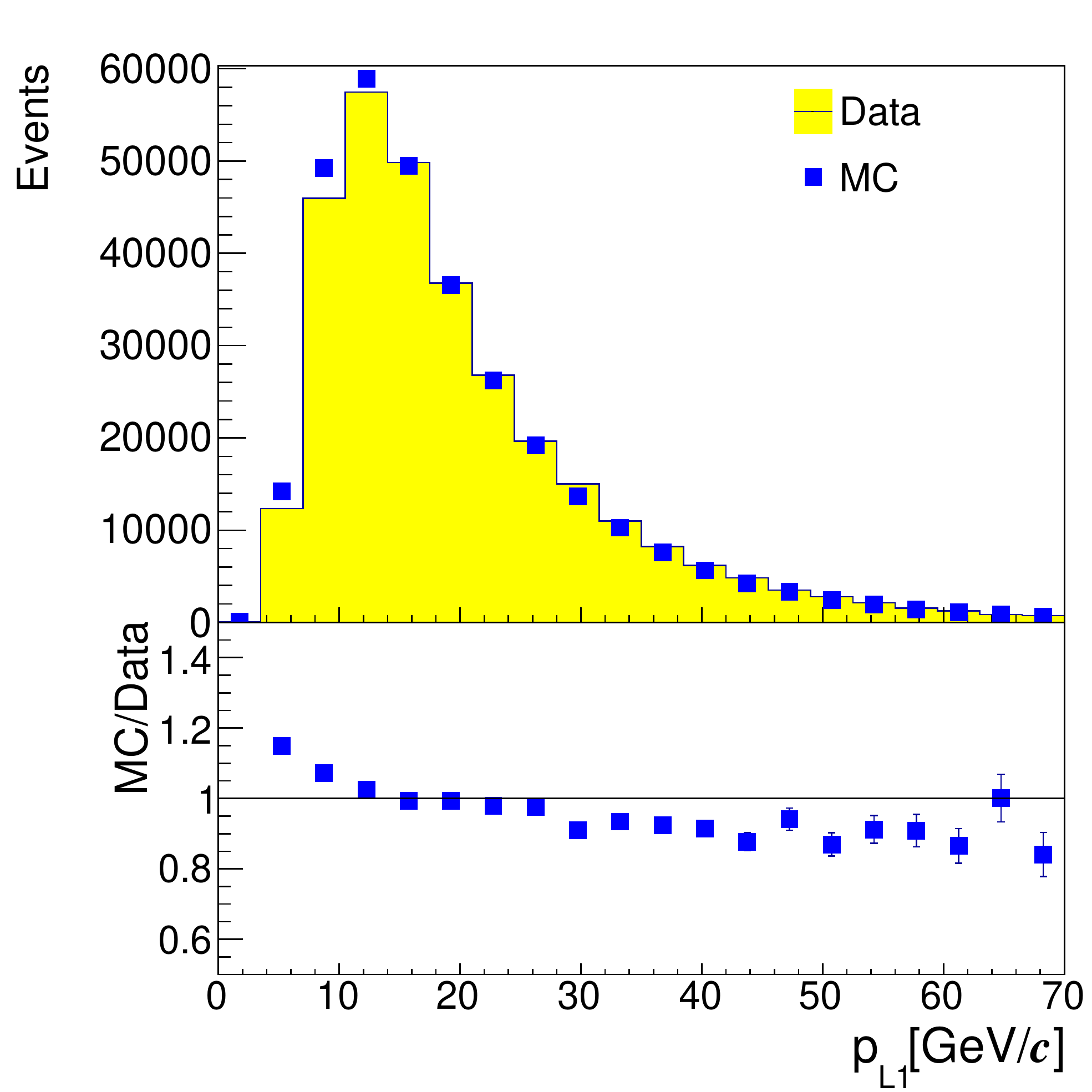}%
            &\includegraphics[width=0.32\textwidth]{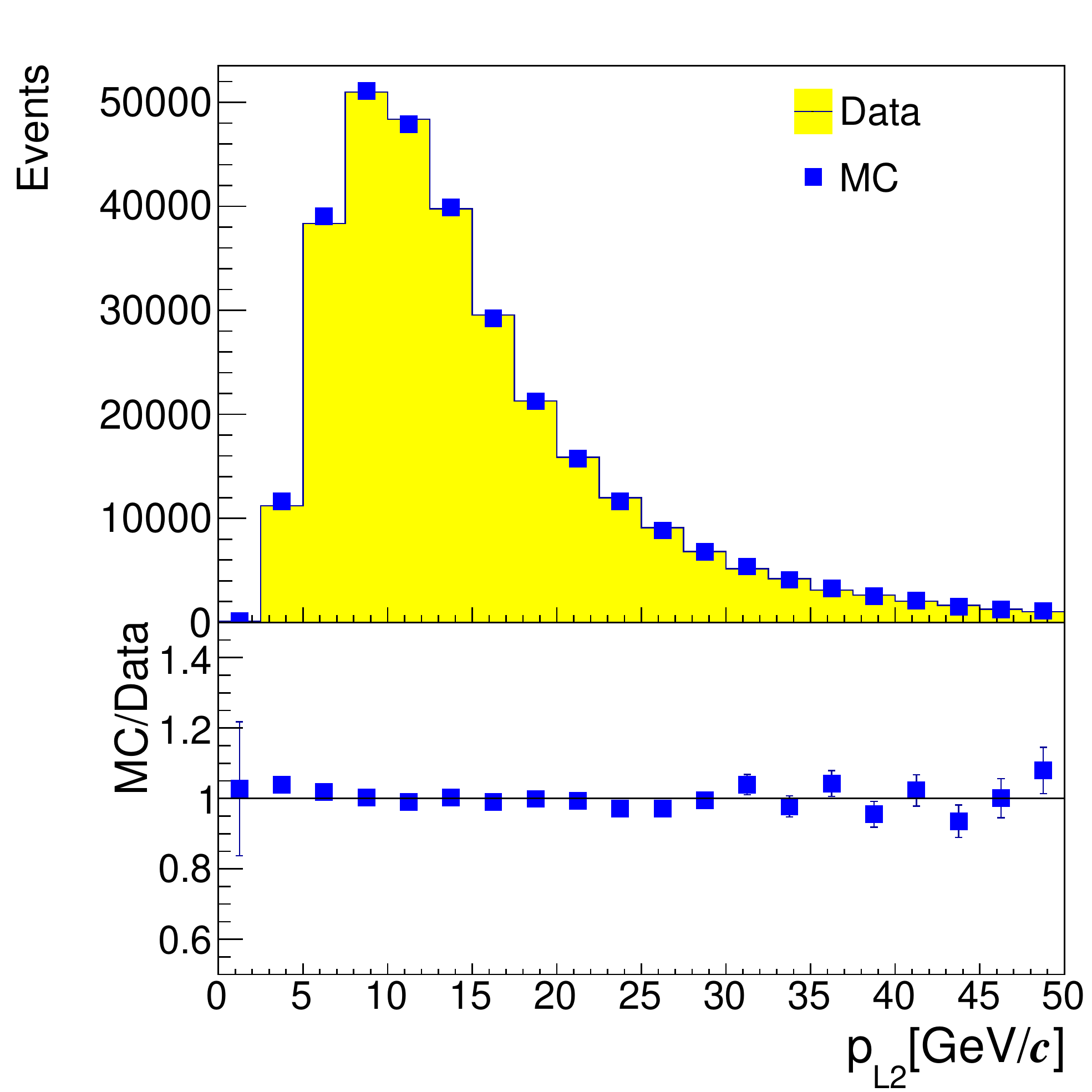}\\%
                \end{tabular}
     \caption{Comparison of distributions of kinematic variables between experimental and MC high-$p_T$ deuteron data.}\label{fig:MCvsData_deuteron}
\end{figure}

\begin{figure}[tbp]
		\begin{tabular}{ccc}			
			\includegraphics[width=0.32\textwidth]{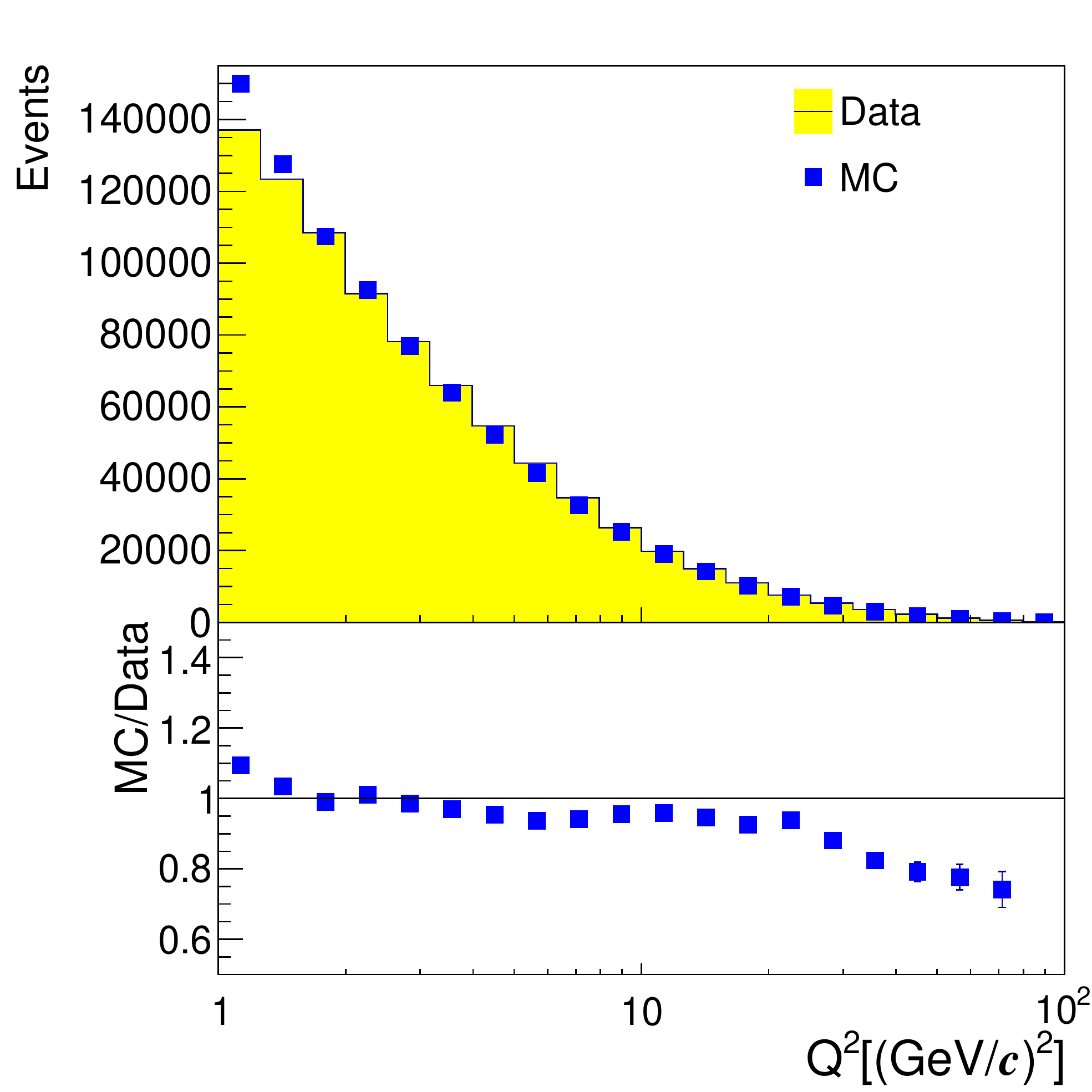}%
			&\includegraphics[width=0.32\textwidth]{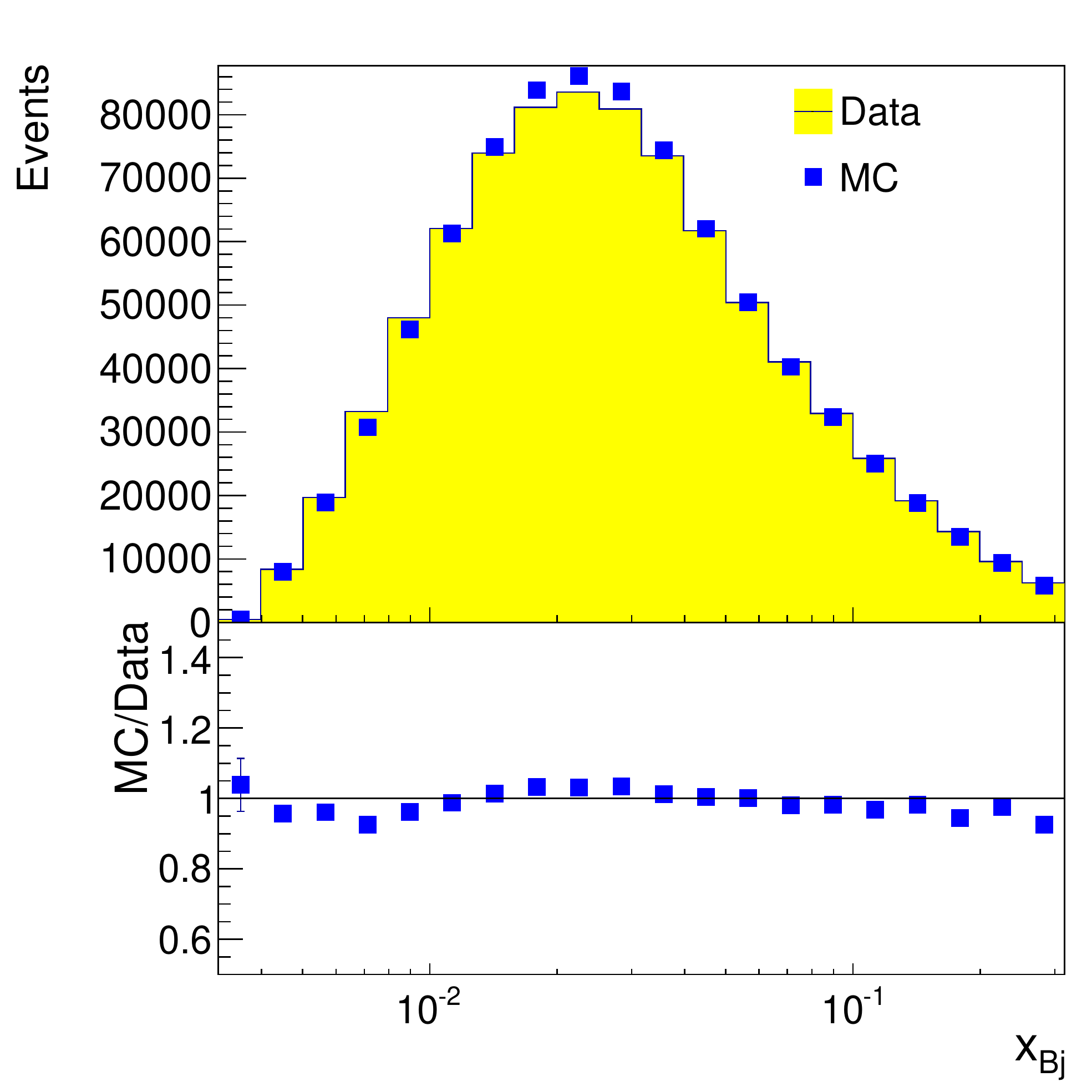}%
			&\includegraphics[width=0.32\textwidth]{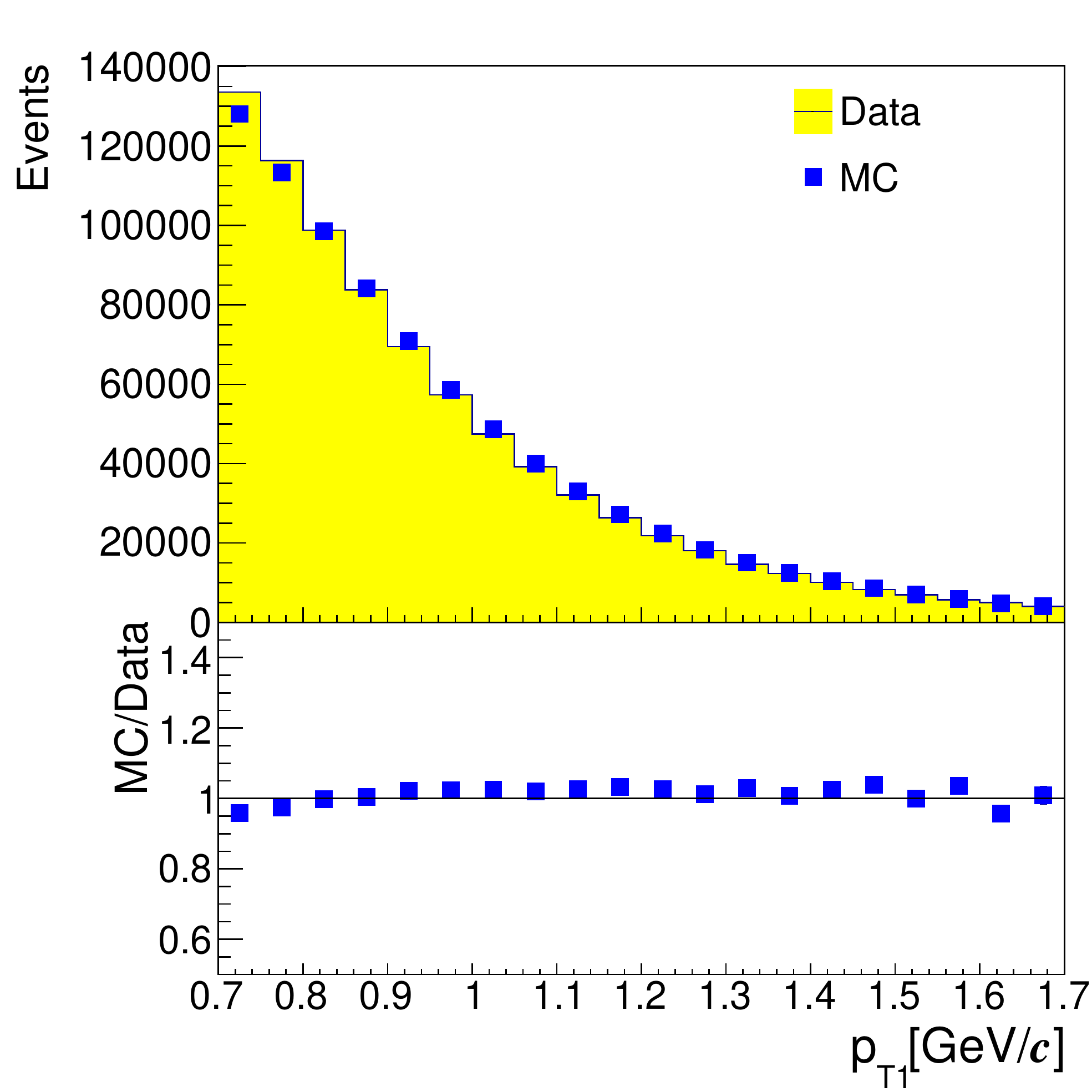}\\
			\includegraphics[width=0.32\textwidth]{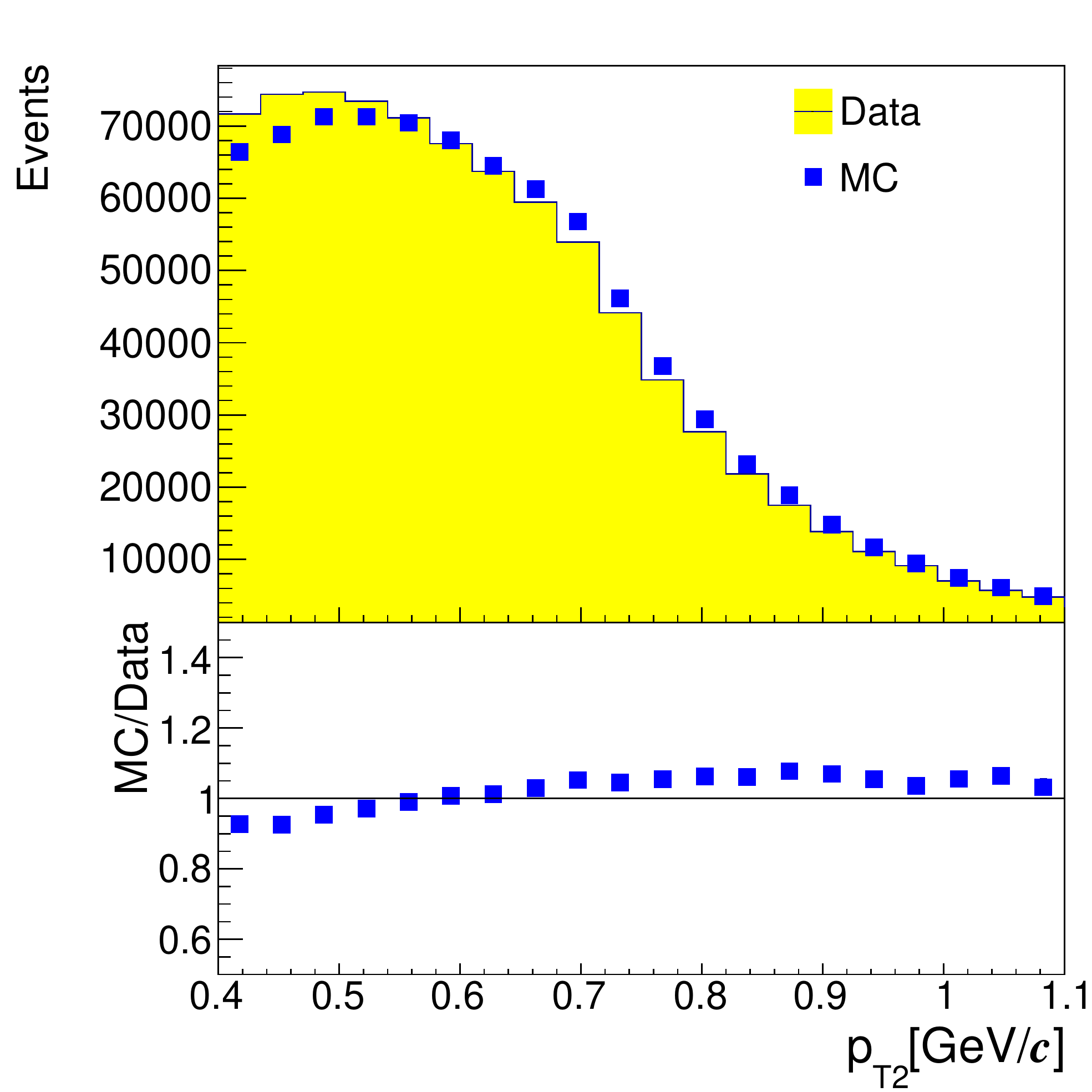}%
            &\includegraphics[width=0.32\textwidth]{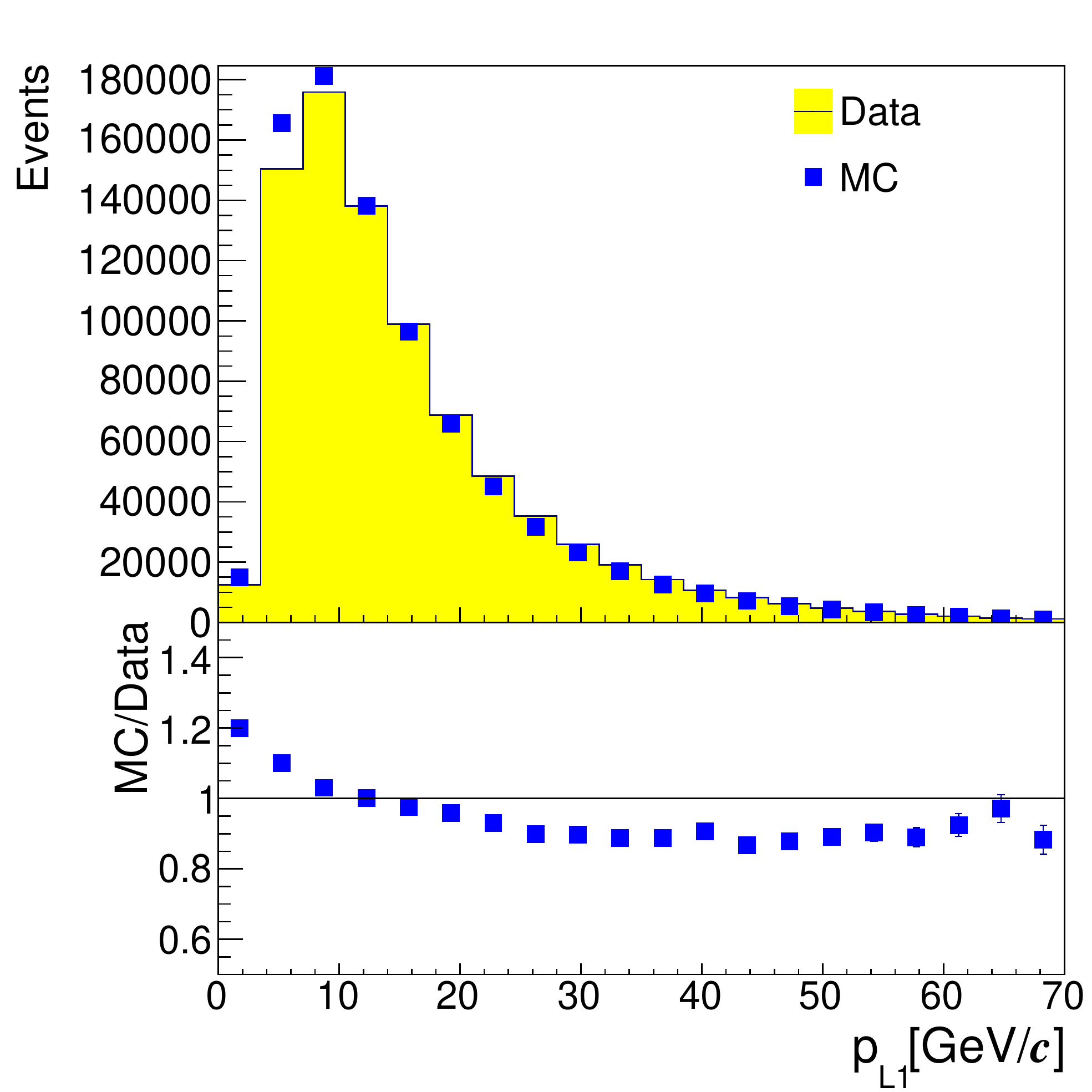}%
            &\includegraphics[width=0.32\textwidth]{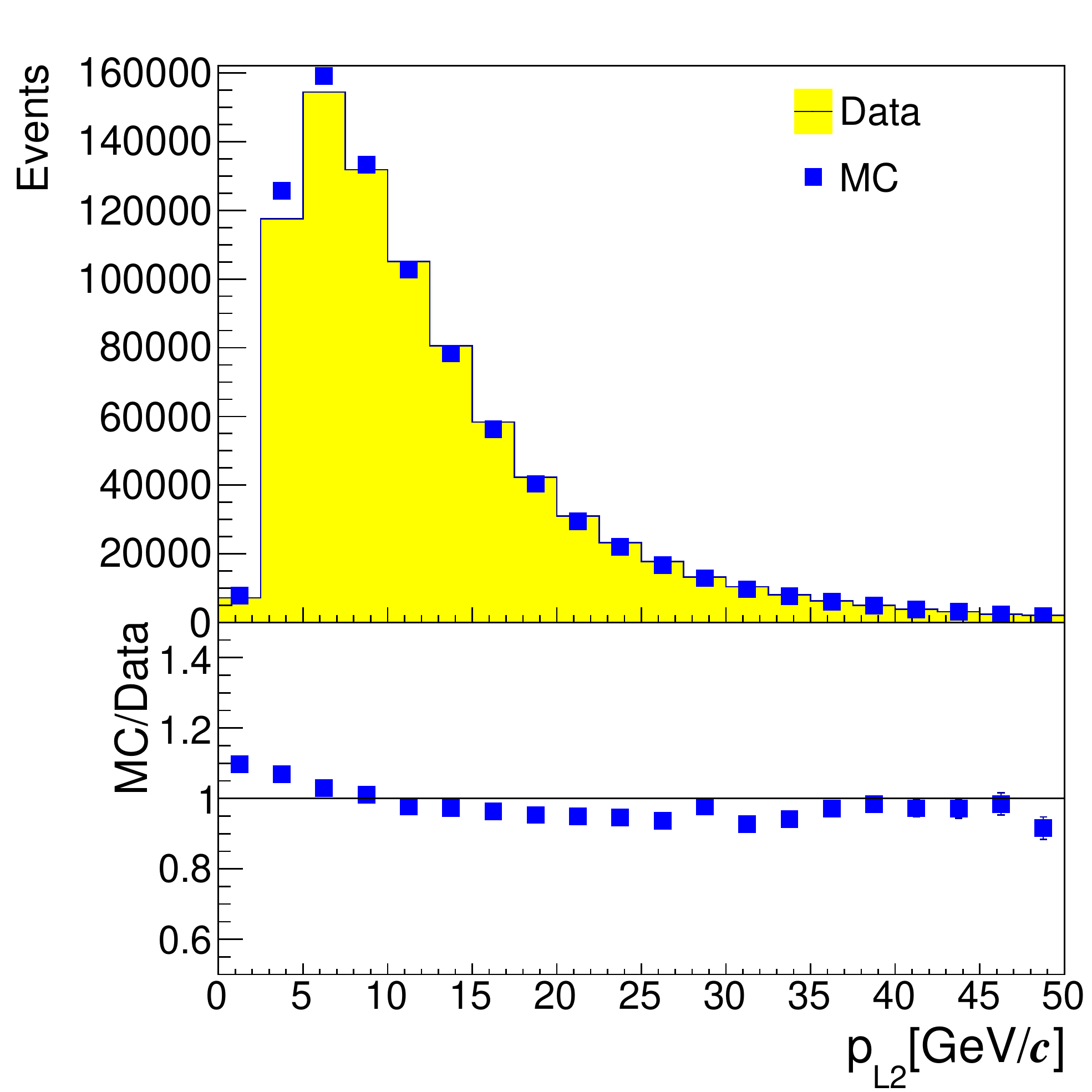}\\%
                \end{tabular}
     \caption{Comparison of distributions of kinematic variables between experimental and MC high-$p_T$ proton data.}\label{fig:MCvsData_proton}
\end{figure}

 
 Using the LEPTO generator (version 6.5)~\cite{Ingelman:1996mq}, two separate MC data samples were produced to simulate the deuteron and proton data. The generator is tuned to the COMPASS data sample obtained with the high-$p_T$ hadron-pair selection as described in Ref.~\cite{Adolph:2012vj}. The MSTW08 parameterisation of input PDFs~\cite{Arneodo:1996qe} was chosen as it gives a good description of the $F_2$ structure function in the COMPASS kinematic range and is valid down to $Q^2=1$~(GeV$/c)^2$. Electromagnetic radiative corrections~\cite{Akhundov:1994my} were applied as a weighting factor to the MC distributions shown in Figures~\ref{fig:MCvsData_deuteron} and~\ref{fig:MCvsData_proton} but not in the MC samples used in NN training. This difference was studied and it was estimated to be negligible.\par 
The generated events were processed by COMGEANT, the COMPASS detector simulation program based on GEANT3. The MC samples for the proton and deuteron data differ in the target material and in the spectrometer set-up. The FLUKA package~\cite{FLUKA} is used in order to simulate secondary interactions. As the next step, the COMPASS reconstruction program CORAL was applied. The same data selection as for real events was used for MC data. Figures~\ref{fig:MCvsData_deuteron} and ~\ref{fig:MCvsData_proton} show the comparison between experimental and MC data for the case of the deuteron and proton data, respectively.
\par
\begin{figure}[tbp]
\centering
\includegraphics[width=0.5\textwidth]{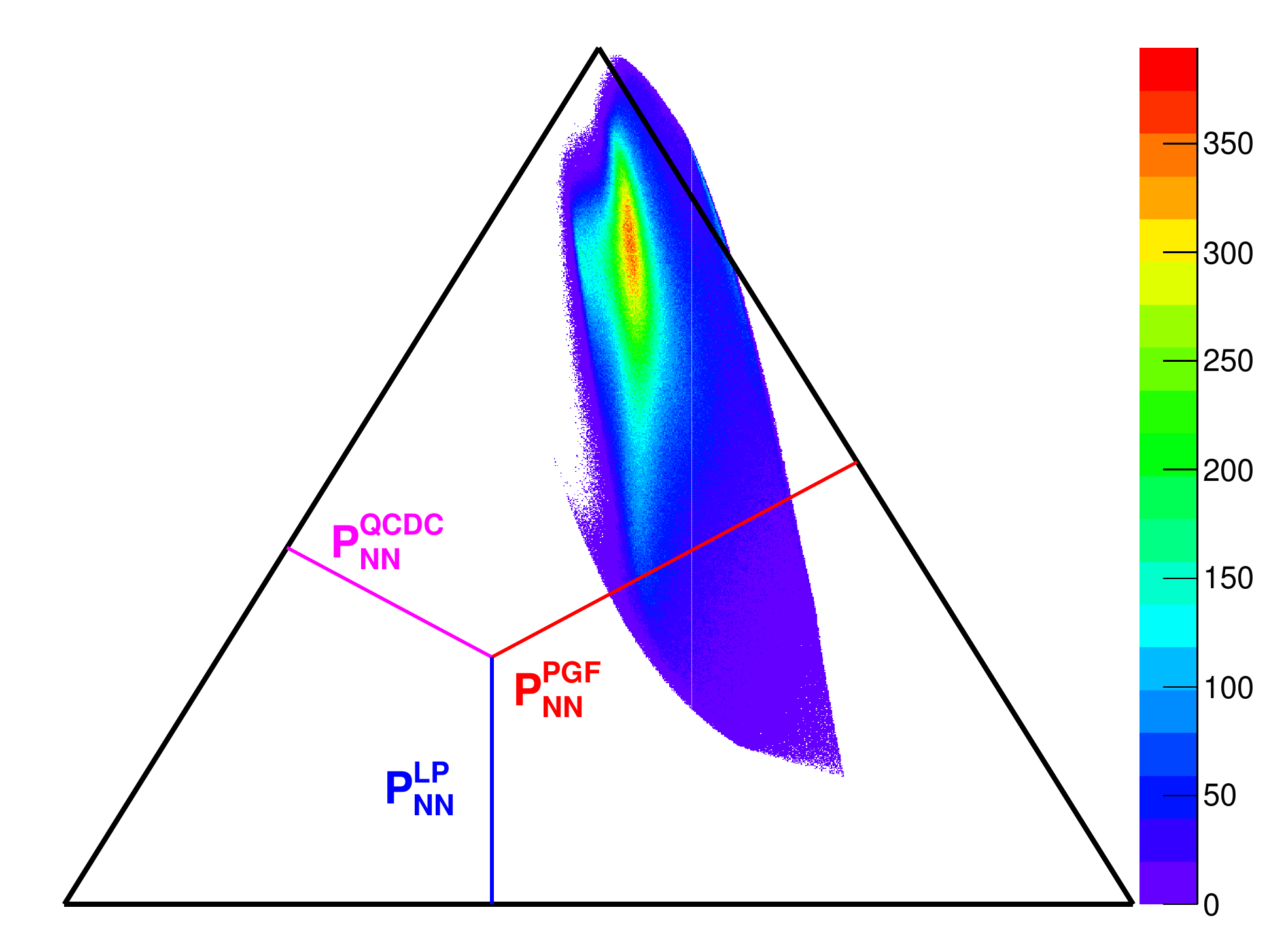}
\caption{  Neural network output after training the MC sample for the proton analysis. The probability of each process for a point representing an event is proportional to the distance to the opposite side of the triangle as in the ``Mandelstam representation''. An exemplary set of probabilities assigned to a point is shown by the colour lines.}
\label{fig:NN_Triangle}
\end{figure}

The main goal of the NN parameterisation is the estimation of the process fractions $R_j$. In the typical case of signal and background separation, the expected NN output would be set to one for the signal and zero for the background. The output value returned by the NN would then correspond to the fraction of signal events in the sample in the given phase space point of the input parameter vector. In the present analysis, the process fractions were estimated simultaneously. In order to have a closure relation on the process probabilities, the sum of them must add up to one, hence only two independent output variables from the NN are needed.
 The estimation of the process fractions $R^j$ from the NN output is accomplished by assigning to each event the probabilities $P^{\text{PGF}}_{\text{NN}}$, $P^{\text{QCDC}}_{\text{NN}}$ and $P^{\text{LP}}_{\text{NN}}$.
 The distribution of the NN output after training is shown in Fig.~\ref{fig:NN_Triangle} on the ``Mandelstam representation'', i.e. as points in an equilateral triangle
with unit height. Points outside of the triangle refer to one estimator being negative, which is possible because in the training the estimators are not bound to be positive. The direct separation of the PGF process using this distribution is statistically less efficient than weighting each event by the three probabilities obtained from the NN output values. These probabilities are used as  values of the process fractions ($R_{\text{PGF}}$, $R_{\text{QCDC}}$ and $R_{\text{LP}}$)  in the data analysis described in Section~\ref{sect:weighted_method}.\par
\begin{figure}[tbp]
        \centering
			\includegraphics[width=0.7\textwidth]{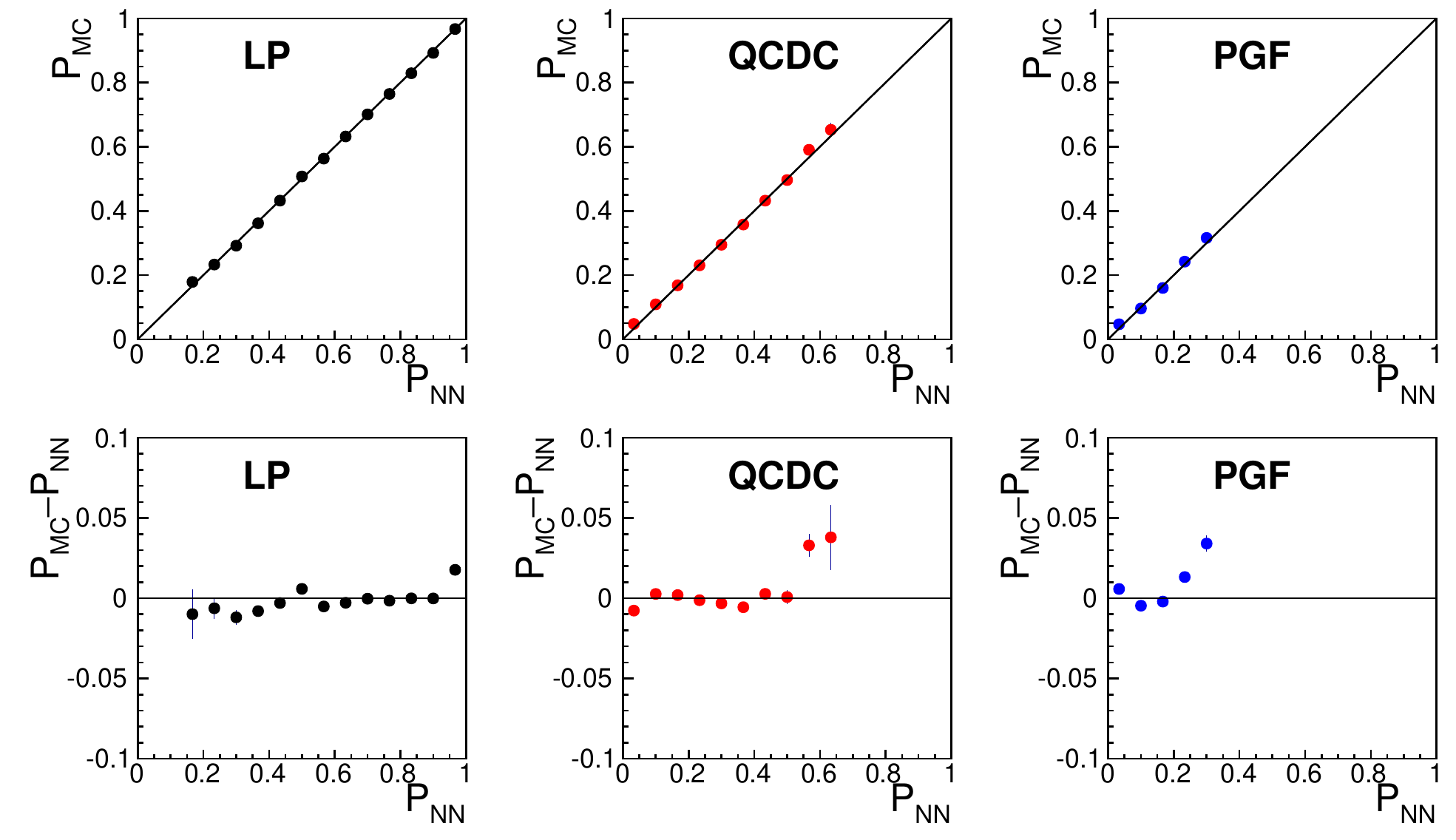}
        \caption {Top panels: Neural network validation. Here $P_{\text{NN}}$ is the fraction of the process given by the NN and $P_{\text{MC}}$ is the true
        fraction of each process from MC in a given $P_{\text{NN}}$ bin. Bottom panels: Difference $P_{\text{MC}}-P_{\text{NN}}$ per bin.} \label{fig:NN_validation}
    \end{figure}

For the validation of the NN training, a statistically independent MC sample is used to check how the NN works on a sample different from the one used for the training. In each bin of $P_{\text{NN}}$ (the value assigned to every MC event by the trained NN), the true fraction  obtained from LEPTO based on the process ID, $P_{\text{MC}}$, is calculated. The results for the NN trained with a MC sample for the proton data are presented in Fig.~\ref{fig:NN_validation}. 
Altogether, the agreement between $P_{\text{NN}}$ and $P_{\text{MC}}$ for all three processes is satisfactory. However, for the two last bins of PGF, the two last bins of QCDC and the last bin of LP the neural network output does not coincide with the true fraction of the given process. This discrepancy concerns a small part of the event sample and is included as a part of the MC-dependent systematic uncertainty. Because of the good agreement between MC and real data, it is assumed that the fractions in Eq.~\ref{eq:decomposition} can be taken from the trained NN, which means that on average $R_j=P_{\text{NN}}^j$.

\section{ Systematic uncertainties}
\label{sect:systematics}
The main source of systematic uncertainties is the dependence of the final results on the Monte Carlo settings and tuning. In order to estimate this uncertainty, different MC settings were used in the process of neural network training. Different combinations of fragmentation parameters were used, the default LEPTO tuning or the COMPASS tuning for the high-$p_T$ selected sample. The event generation was done with and without the `Parton Shower’~\cite{Bengtsson:1987rw}. Two PDF sets were used (MSTW08 or CTEQ5L~\cite{Lai:1999wy}). Two different parameterisations of the longitudinal structure function $F_L$ are used, either from LEPTO or the $R=\sigma_L/\sigma_T$ parameterisation of Ref.~\cite{Abe:1998ym}. For secondary interactions, either the FLUKA or the GHEISHA~\cite{Fesefeldt:1985yw} package were used. \par
\begin{figure}[tbp]
    \centering\includegraphics[width=0.75\textwidth]{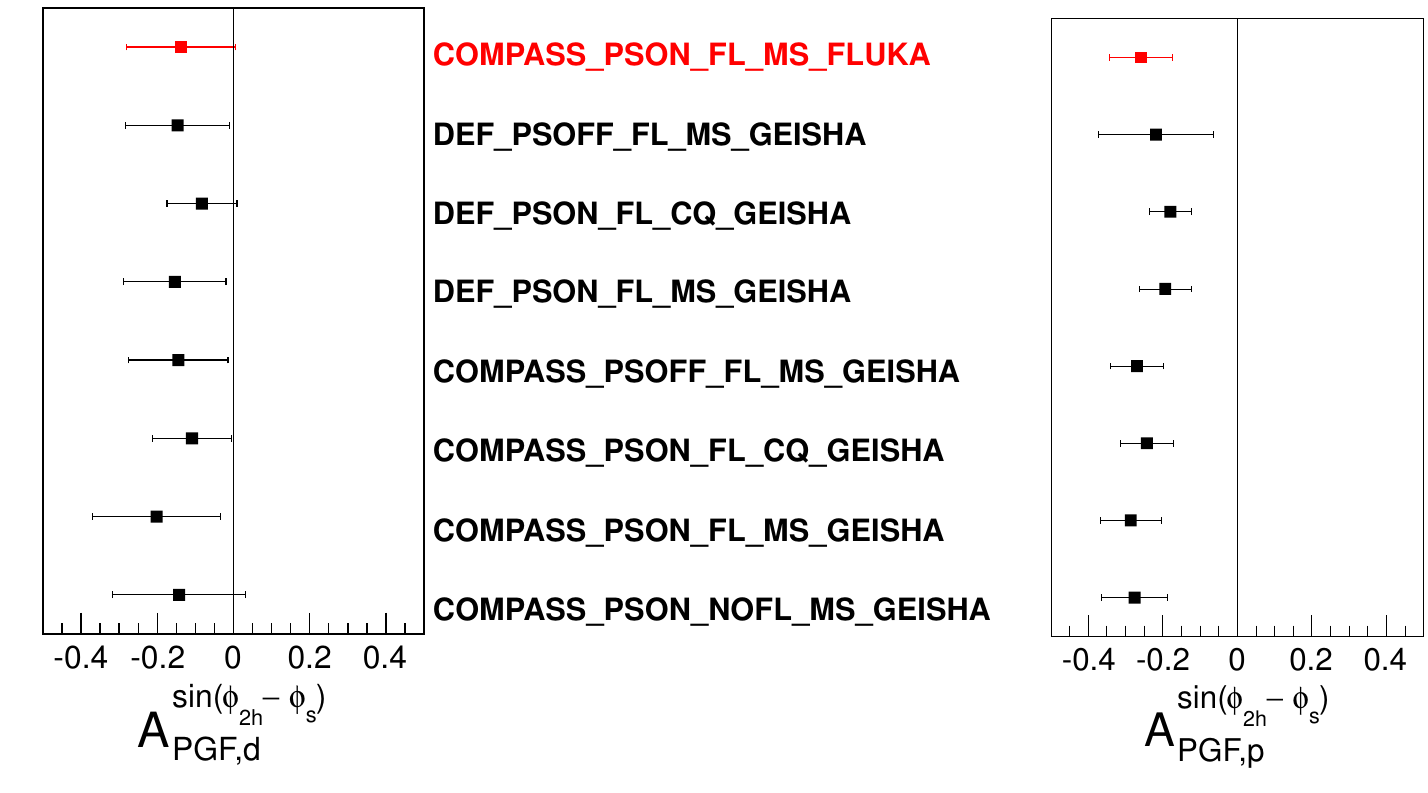}
    \caption{Systematic changes in the final result caused by using different MC settings. Besides the final result shown on the top, seven other results are shown that are obtained with MC samples that differ by the choice of COMPASS or default LEPTO tuning, `Parton Shower’ on or off, $F_L$ from LEPTO or from $R=\sigma_L/\sigma_T$, MSTW or CTEQ5L PDF sets, FLUKA or GHEISHA for secondary interactions. The results for deuteron data are shown in the left panel and for the proton data on the right panel.}
\label{fig:MC_systematics}
\end{figure}

Figure~\ref{fig:MC_systematics} shows the results for the gluon Sivers asymmetry when using eight different MC productions of the deuteron and proton data. The final result is presented on the top. These two MC productions, using FLUKA and Parton Shower, yield the best comparison between experimental and MC data, which is shown in Fig.~\ref{fig:MCvsData_deuteron} and \ref{fig:MCvsData_proton}. The systematic uncertainty originating from different MC tunings is calculated as $(A^{\text{PGF}}_{max}-A^{\text{PGF}}_{min})/2$.\\
The systematic uncertainty due to false asymmetries was studied by extracting the asymmetries between the two parts of the same target cell. The results are found to be compatible with zero. Furthermore it was checked how a small  artificial false Sivers asymmetry influences the final result. When a false asymmetry of 1\% is introduced, for both proton and deuteron data the final result changes by 25\% of the statistical error. No systematic uncertainty is assigned to account for false asymmetries.\par
The final state of the photon-gluon-fusion process is a quark-antiquark pair. Thus most of the hadron pairs produced from this subprocess should have opposite charge. Although a selection $q_1q_2=-1$ slightly increases the ($\phi_g$, $\phi_{P}$) correlation, it also reduces the statistics. The results with and without this requirement are statistically consistent. The requirement of opposite charges of the two hadrons is hence not included in the data selection.\par
Radiative corrections were not included in the MC production that is used in the main analysis of this letter. In order to estimate the systematic uncertainty introduced by this omission, a separate MC sample is used that was produced for the 2006 COMPASS set-up including radiative corrections based on RADGEN~\cite{Akushevich:1998ft}. The difference in the final value for the gluon Sivers asymmetry for the proton is only $0.018$, which corresponds to $21\%$ of the statistical uncertainty. A corresponding systematic uncertainty is assigned due to the fact that radiative corrections are not included in the MC simulations and hence in the NN training.\par
Our results are obtained in only one $x_g$ bin for $A^{\text{Siv}}_{\text{PGF}}$, one $x_C$ bin for $A^{\text{Siv}}_{\text{QCDC}}$ and one $x_{Bj}$ bin for $A^{\text{Siv}}_{\text{LP}}$. As the asymmetries are strongly correlated binning in $x_{Bj}$ affects the values of $A^{\text{Siv}}_{\text{PGF}}$ and $A^{\text{Siv}}_{\text{QCDC}}$, which are extracted in a single bin as before. The $A^{\text{Siv}}_{\text{PGF}}$ result changes by $0.07$ for deuteron data and $0.011$ for proton data when two $x_{Bj}$ bins are introduced, and these values are taken as an estimate of the related systematic uncertainty (see Table~\ref{tab:systematics}).\par
The asymmetries $A^{\text{Siv}}_j$ of Eq.~\eqref{eq:events_number} were also extracted using the unbinned maximum likelihood  method that, as expected, yields the same results of $A_{\text{PGF}}$ as the above described analysis. Concerning the orthogonality of different modulations of the cross section, it was checked by what amount the Sivers asymmetry changes, when also the other seven asymmetries were included in the fit (see above). The change in the final result of the PGF asymmetry is negligible for both deuteron and proton data.

 The systematic uncertainties on target polarisation and dilution factor are multiplicative and estimated to be about  $5\%$ and $2\%$ of the statistical uncertainty, respectively. The final systematic uncertainty is obtained by summing all components in quadrature. All above mentioned contributions and the final systematic uncertainty are listed in Table~\ref{tab:systematics}.
\begin {table} [htpb]
\centering
\caption{Summary on systematic uncertainties of the final values of the gluon Sivers asymmetry for deuteron and proton data.}
\label{tab:systematics}
\begin {tabular}{|c|c|c|c|c|} 
\hline

 &\multicolumn{2}{c}{deuteron data}& \multicolumn{2}{|c|}{proton data}\\\hline
source & uncertainty & fraction of $\sigma_{stat}$ & uncertainty & fraction of $\sigma_{stat}$  \\ \hline\hline
Monte Carlo settings&0.060& 40\% & 0.054 & 64\% \\ \hline
radiative corrections &0.018 & 12\% & 0.018& 21\% \\ \hline
one or two $x_{Bj}$ bins & 0.07 & 47\% & 0.011 & 13\% \\ \hline
include 7 other asymmetries &0.003 &  2\% & 0.005 &6\%  \\ \hline
target polarisation & 0.0075 & 5\%  & 0.0043& 5\%  \\ \hline
dilution factor     & 0.003 & 2\%  & 0.0018& 2\% \\ \hline \hline
total $\sqrt{\sum \sigma_i^2}$ &0.10 & 63\% & 0.06 & 69\% \\\hline
\end{tabular}
\end{table}

\section{Results}
\label{sect:results}
\begin{figure}[tbp]
  \begin{tabular}{cc}
     \includegraphics[width=0.35\paperwidth]{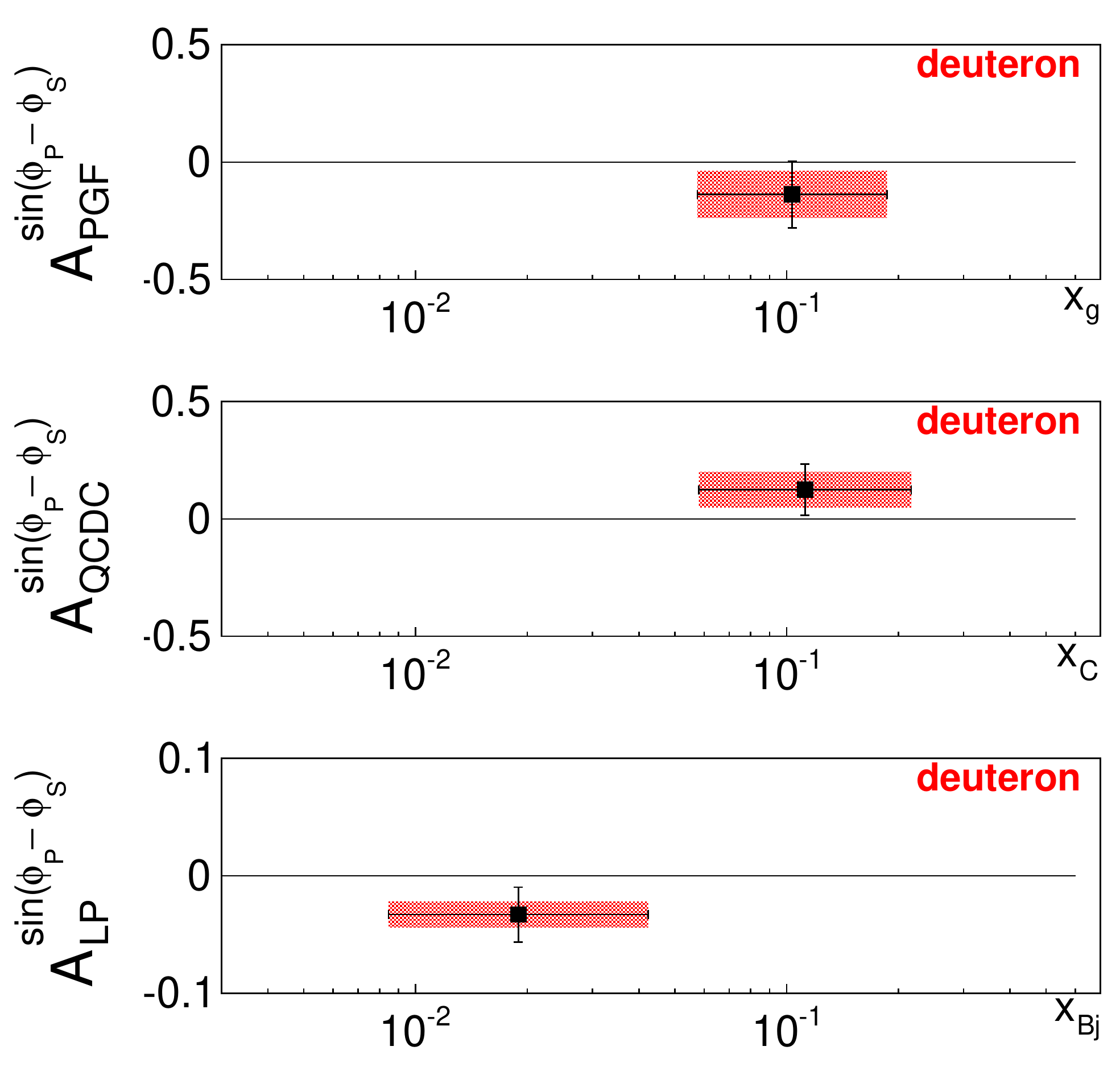}&
     ~~~~\includegraphics[width=0.35\paperwidth]{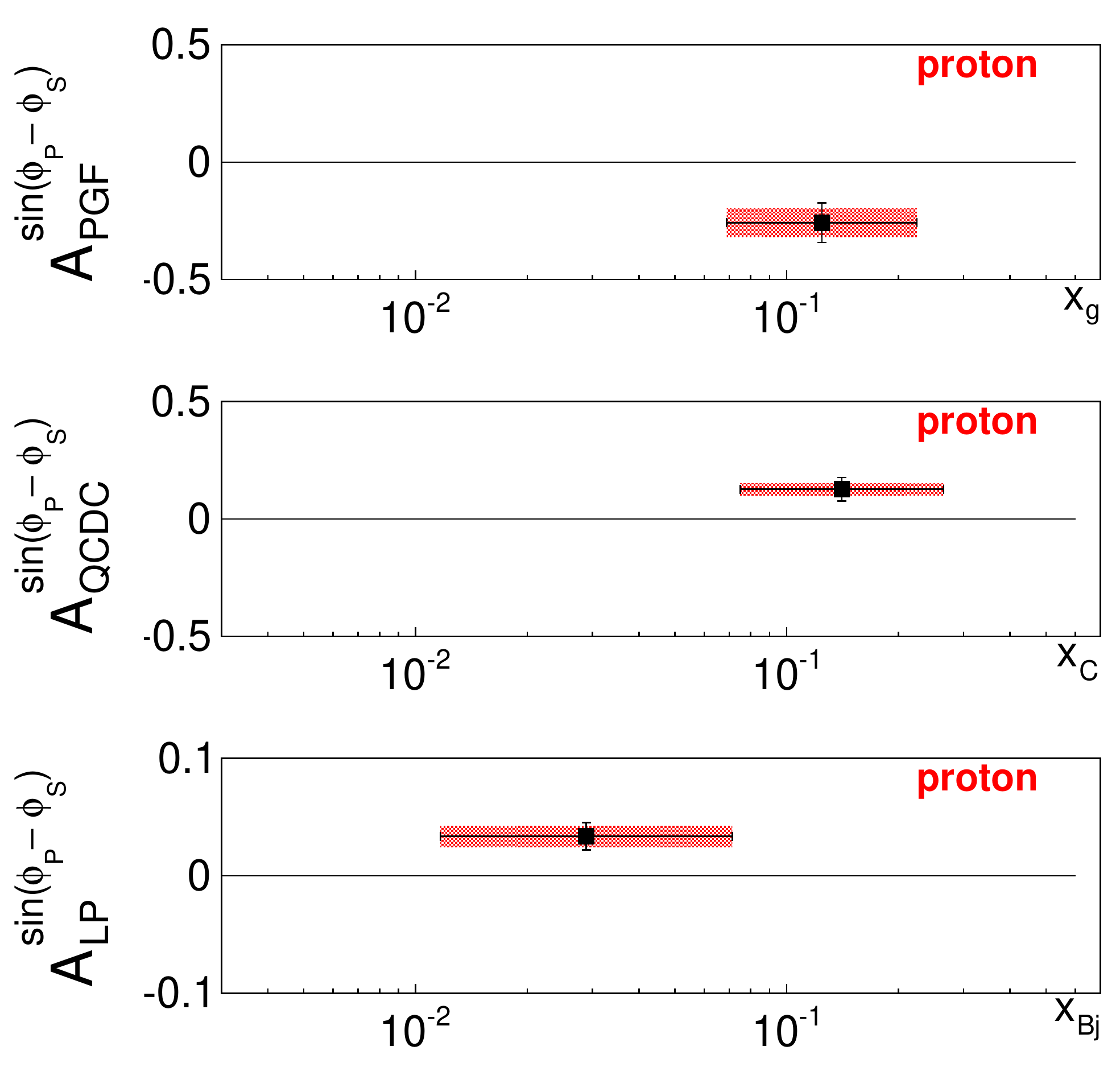}
    \end{tabular}
    \caption{ Sivers two-hadron asymmetry extracted for Photon-Gluon fusion (PGF), QCD Compton (QCDC) and Leading Process (LP) from the COMPASS high-$p_T$ deuteron (left) and proton (right) data. The $x$ range is  the RMS of the logarithmic distribution of $x$ in the MC simulation. The red bands indicate the systematic uncertainties. Note the different ordinate scale used in the third row of panels.}\label{fig:results}
\end{figure}

The method presented in Section~\ref{sect:weighted_method} with the use of trained neural networks was applied to the two data sets described in Section~\ref{sect:setup}. The gluon Sivers asymmetry as extracted from lepton nucleon DIS, in which at least two high-$p_T$ hadrons are detected, is shown in Fig.~\ref{fig:results} and presented in Table~\ref{tab:results} together with the contribution of the two other hard processes, {\it i.e.} QCD Compton and leading process. The result of the analysis of the deuteron data is $A^{Siv,d}_{PGF} = -0.14 \pm 0.15 (\text{stat.})\pm0.10 (\text{syst.})$ measured at ~$\langle x_g\rangle = 0.13$. The proton result, $A^{Siv,p}_{PGF} = -0.26 \pm 0.09 (\text{stat.})\pm 0.06 (\text{syst.})$  obtained at ~$\langle x_g\rangle = 0.15$, is consistent with the deuteron result within less than one standard deviation of the combined statistical uncertainty. The two results are expected to be consistent, as presumably the transverse motion of gluons is the same in neutron and proton. Combining the proton and deuteron results, the measured effect is negative, $A_{PGF}^{Siv}=-0.23\pm 0.08(\text{stat})\pm0.05(\text{syst})$, which is away from zero by more than two standard deviations of the quadratically combined uncertainty. This result is particularly interesting in view of the gluon contribution to the proton spin. A non-zero gluon Sivers effect is a signature of gluon transverse motion in the proton~\cite{Burkardt:2002ks}. The recent analysis of the PHENIX data~\cite{D'Alesio:2015uta}
gives a gluon Sivers effect for protons, which is compatible with zero. The COMPASS result for the proton target is
negative and more than two standard deviations below zero, but it should be noted that the two
results are obtained for different centre of mass energy and $x_g$ values. Even more important,
one has to recall that
the existence of colour gauge links complicates the picture, as they lead to two different universal gluon
Sivers functions, which in the different processes combine with process-dependent calculable factors~\cite{Buffing:2013kca}.
As a result, the gluon Sivers function that appears in one process can be different from the one appearing
in a different process, and assessment of compatibility requires a deeper theoretical analysis. 
\par

\par
\begin{table}[H]
\centering
\caption{Summary of Sivers asymmetries, $A^{Siv}_{PGF}, A^{Siv}_{QCDC}, A^{Siv}_{LP}$, obtained for deuteron and proton data.}
\label{tab:results}
\scalebox{0.8}{
\begin {tabular}{|c|c|c|c|c|c|c|} 
\hline
 &\multicolumn{3}{c}{deuteron data}& \multicolumn{3}{|c|}{proton data}\\\hline
process & asymmetry & statistical error  & systematic uncertainty & asymmetry & statistical error  & systematic uncertainty    \\ \hline\hline
PGF & -0.14 & 0.15 & 0.10 & -0.26 & 0.09 & 0.06 \\ \hline
QCDC & 0.12 & 0.11 & 0.08 & 0.13 & 0.05 & 0.03 \\ \hline
LP & -0.03 & 0.02 & 0.01 & 0.03 & 0.01 & 0.01 \\ \hline
\end{tabular}
}
\end{table}

For the asymmetry of the leading process, the high-$p_T$ sample of the COMPASS proton data has provided a positive value (see Fig.~\ref{fig:results} right-bottom panel). It can be compared with the COMPASS results on the Sivers asymmetry for charged hadrons produced in SIDIS $\ell p\rightarrow\ell' h^{\pm}X$ single-hadron production~\cite{Adolph:2012sp}, which for negative hadrons was found to be about zero and for positive hadrons different from zero and positive, so that for the two-hadron final state a positive value may indeed be expected.\par
\begin{figure}[tbp]
  \begin{tabular}{cc}
     \includegraphics[width=0.35\paperwidth]{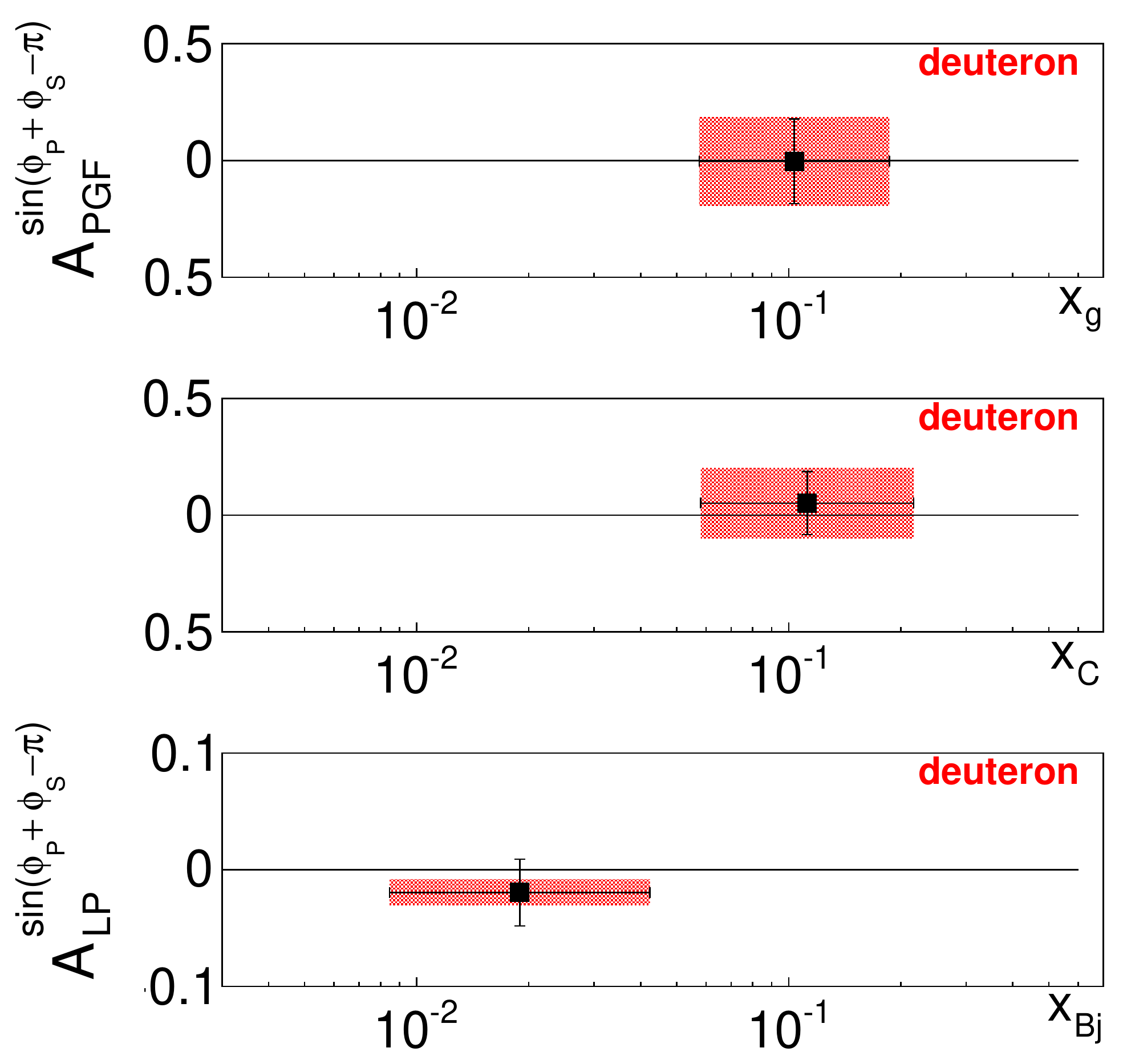}&
     ~~~~\includegraphics[width=0.35\paperwidth]{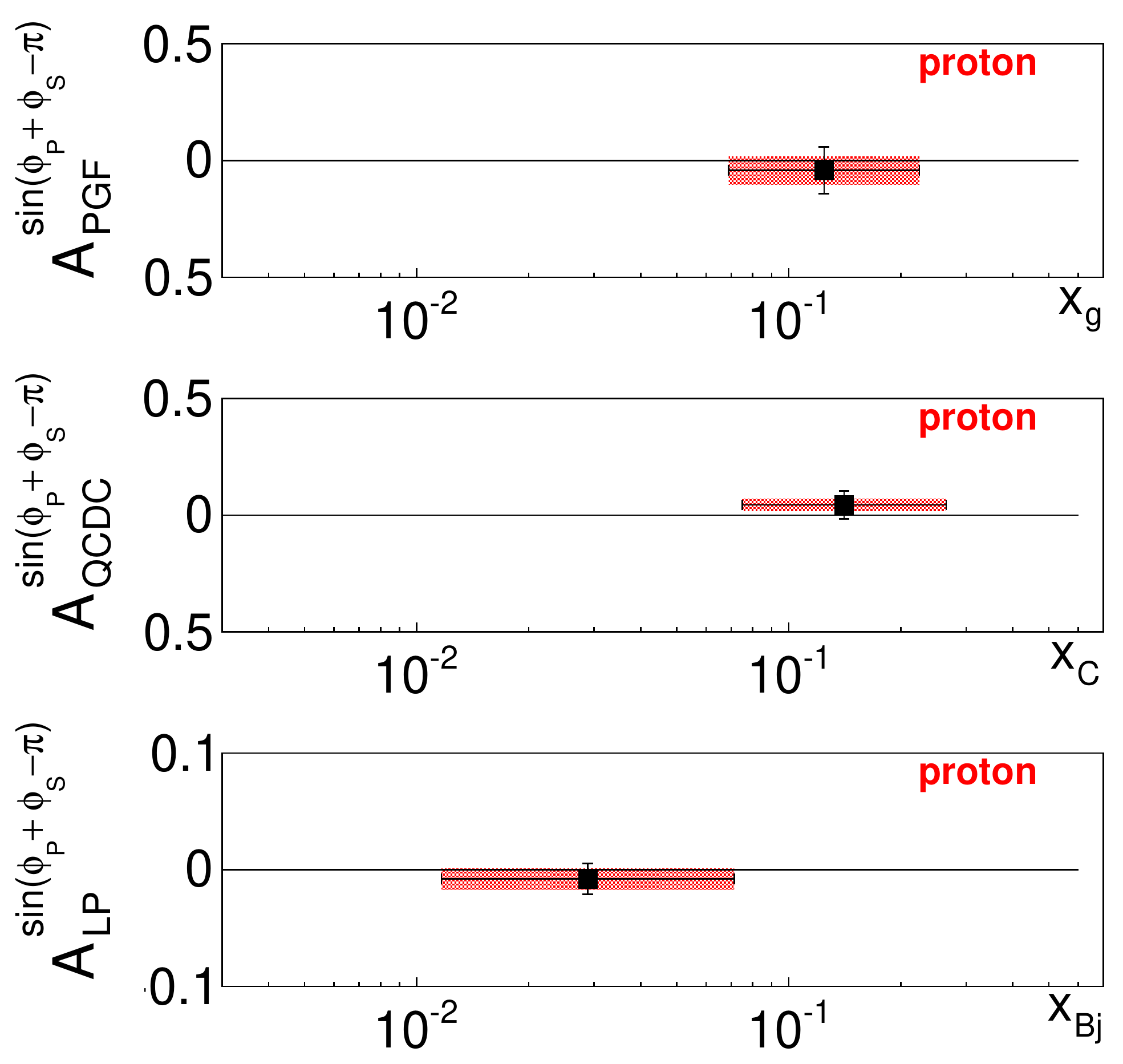}
    \end{tabular}
    \caption{Collins-like two-hadron asymmetry extracted for Photon-Gluon fusion (PGF), QCD Compton (QCDC) and Leading Process (LP) from the COMPASS high-$p_T$ deuteron (left) and proton (right) data. The $x$ range is  the RMS of the logarithmic distribution of $x$ in the MC simulation. The red bands indicate the systematic uncertainties.}
\label{fig:collins_results}
\end{figure}

The same analysis method was also applied to extract the Collins-like asymmetry for charged hadrons, {\it i.e.} the cross section dependence on the sine of the Collins angle $(\phi_P+\phi_S-\pi)$. To this purpose, the asymmetries $A^{\sin{(\phi_P+\phi_S-\pi)}}_{\text{PGF}}$, $A^{\sin{(\phi_P+\phi_S-\pi)}}_{\text{QCDC}}$, $A^{\sin{(\phi_P+\phi_S-\pi)}}_{\text{LP}}$ were determined for the same COMPASS
high-$p_T$ deuteron and proton data samples. The results are shown in Fig.~\ref{fig:collins_results}. The amplitude of the Collins
modulation for gluons is found to be consistent with zero, in agreement with the naive expectation that is
based on the fact that there is no gluon transversity distribution~\cite{Barone:2001sp}. Recently it was suggested that
a transversity-like TMD gluon distribution $h_1^g$ could generate a $\sin{(\phi_S + \phi_T)}$ modulation
in leptoproduction of two jets or heavy quarks~\cite{Boer:2016fqd}. In this case the results shown in Fig.~\ref{fig:collins_results} provide a bound to the
size of $h_1^g$. The results given in the present letter can also be interpreted such that no  false
systematic asymmetry is introduced by the rather complex analysis method used, and that the
result obtained for the gluon Sivers asymmetry, which is definitely different from zero, is strengthened.
It should also be noted that the Collins-like asymmetry of the leading process for the proton
is found to be consistent with zero for high-$p_T$ hadron pairs, in qualitative agreement with the measurement
of the Collins asymmetry in single-hadron SIDIS measurement~\cite{Adolph:2012sn}, where opposite values of about equal
size were observed for positive and negative hadrons.

\par
\section{Summary and conclusions}
The Sivers asymmetry for gluons is extracted from the measurement of high-$p_T$ hadron pairs in SIDIS at COMPASS off transversely polarised deuterons and protons.
The analysis is very similar to the one already used by the COMPASS collaboration in
order to measure $\Delta g/g$ , the gluon polarisation in a longitudinally polarised
nucleon. The large kinematic acceptance and the high energy of the muon beam
make the sample containing two high-$p_T$ hadrons sufficiently large for the
present analysis, which is limited to a small part of the accessible phase space. The criteria applied to select hadron pairs allow to enhance the contribution of the photon-gluon fusion process with respect to the leading-order virtual-photon absorption process. The Sivers asymmetry was then obtained as the
amplitude of the sine modulation in the Sivers angle, $\phi_{Siv} = \phi_P - \phi_S$.

In spite of the enrichment of the PGF  fraction in the high-$p_T$ hadron pair sample,
in order to single out the contribution of the PGF process to the asymmetry it is necessary to subtract the
contributions from the other two processes, LP and QCDC. In this analysis, the fractions of the three processes
were determined from MC algorithms, and the three corresponding asymmetries
were extracted from the data using a NN technique. Therefore, the analysis requires a
precise MC description of the data, so that these quantities can be calculated reliably.
Since the results derived from a NN approach strongly depend on the Monte Carlo sample on which
the network was trained, much effort was devoted to obtain a good description of the experimental
data by MC simulations, and the analysis was repeated using eight different MCs and the (small)
differences in the results were included in the systematic uncertainties.

Averaging results obtained from the deuteron and proton data, the measured gluon Sivers asymmetry turns out to be $-0.23\pm0.08(\text{stat.})\pm0.05(\text{syst.})$, which is away from zero by more than two standard deviations of the total experimental uncertainty. This result supports the existence of a transverse motion of gluons in a transversely polarised nucleon, although the quantification of this motion is model-dependent. 

In addition, another result obtained in this work from the same data is the extraction of the Collins-like gluon asymmetry, {\it i.e.} the amplitude of the 
sine modulation of the Collins angle  $\phi_{Col} = \phi_P + \phi_S-\pi$. Recent developments
have hypothesised a non-zero Collins-like gluon asymmetry that however is not related to transversity. Our result on the Collins-like asymmetry, which is obtained from the same hadron-pair data that we used to extract the non-zero result on the gluon Sivers asymmetry, is found to be compatible with zero. 

\section*{Acknowledgements} 

This work was made possible thanks to the financial support of our funding agencies. We also acknowledge the support of the CERN management and staff, as well as the skills and efforts of the technicians of the collaborating institutes. 



\end{document}